\documentclass[journal=cmatex,manuscript=article]{achemso}
\usepackage[version=3]{mhchem} 
\usepackage{pslatex}
\usepackage{xr}
\usepackage{gensymb} 
\usepackage{textcomp} 

\graphicspath{{./Figures/}} 

\author{Christopher L. Rom}
\affiliation{Materials Science Center, National Renewable Energy Laboratory, Golden, CO 80401, USA}
\alsoaffiliation[Colorado State University]
{Department of Chemistry, Colorado State University, Fort Collins, CO, 80523, USA}
\author{Rebecca W. Smaha}
\affiliation{Materials Science Center, National Renewable Energy Laboratory, Golden, CO 80401, USA}
\author{Callan A. Knebel}
\affiliation[Colorado State University]
{Department of Chemistry, Colorado State University, Fort Collins, CO, 80523, USA}
\author{Karen N. Heinselman}
\affiliation{Materials Science Center, National Renewable Energy Laboratory, Golden, CO 80401, USA}
\author{James R. Neilson}
\affiliation[Colorado State University]
{Department of Chemistry, Colorado State University, Fort Collins, CO, 80523, USA}
\alsoaffiliation{School of Advanced Materials Discovery, Colorado State University, Fort Collins, CO, 80523, USA}
\author{Sage R. Bauers}
\affiliation[National Renewable Energy Laboratory]
{Materials Science Center, National Renewable Energy Laboratory, Golden, CO 80401, USA}
\author{Andriy Zakutayev}
\affiliation[National Renewable Energy Laboratory]
{Materials Science Center, National Renewable Energy Laboratory, Golden, CO 80401, USA}
\email{Andriy.Zakutayev@nrel.gov}
\title{
Bulk and film synthesis pathways to ternary magnesium tungsten nitrides
}

\begin{document}
\begin{abstract}
Bulk solid state synthesis of nitride materials usually leads to thermodynamically stable, cation-ordered crystal structures, whereas thin film synthesis tends to favor disordered, metastable phases. 
This dichotomy is inconvenient both for basic materials discovery, where non-equilibrium thin film synthesis methods can be useful to overcome reaction kinetic barriers, and for practical technology applications where stable ground state structures are sometimes required.
Here, we explore the uncharted Mg-W-N chemical phase space, using rapid thermal annealing to reconcile the differences between thin film (on ambient or heated substrates) and bulk powder syntheses.
Combinatorial co-sputtering synthesis from Mg and W targets in a \ce{N2} environment yielded cation-disordered Mg-W-N phases in the rocksalt ($0.1<$ Mg/(Mg+W) $<0.9$), and hexagonal boron nitride ($0.7<$ Mg/(Mg+W) $<0.9$) structure types. 
In contrast, bulk synthesis produced a cation-ordered polymorph of \ce{MgWN2} that consists of alternating layers of rocksalt-like [\ce{MgN6}] octahedra and nickeline-like [\ce{WN6}] trigonal prisms (denoted ``rocksaline").
Thermodynamic calculations corroborate these observations, showing rocksaline \ce{MgWN2} is stable while other polymorphs are metastable.
We also show that rapid thermal annealing can convert disordered rocksalt films to this cation-ordered polymorph, near the \ce{MgWN2} stoichiometry. 
Electronic structure calculations suggest that this rocksalt-to-rocksaline structural transformation should also drive a metallic-to-semiconductor transformation, but our resistivity measurements were only able to confirm the semiconducting behavior of rocksaline \ce{MgWN2} and rocksalt \ce{Mg3WN4}.
In addition to revealing three new phases (rocksalt \ce{MgWN2} and \ce{Mg3WN4}, hexagonal boron nitride \ce{Mg3WN4}, and rocksaline \ce{MgWN2}), these findings highlight how rapid thermal annealing can control polymorphic transformations, adding a new strategy for exploration of thermodynamic stability in uncharted phase spaces.

\end{abstract}

\section{Introduction} 
Ternary nitrides are an emerging class of ceramic materials with applications in solid-state lighting, electrochemical energy storage, optoelectronics, piezoelectrics, ferroelectrics, and wide-bandgap semiconductor devices.\cite{greenaway2021ternaryReview} 
However, nitrides are underexplored, lagging behind oxides with an order of magnitude fewer scientific publications and known structures.\cite{greenaway2021ternaryReview, zakutayev2016designPhotovoltaicNitrides, sun_map_2019} 
Therefore, exploring chemical phase space to find new ternary nitrides will open avenues to identifying novel materials with intriguing functional properties that may underlie future technologies.
Recent breakthroughs in high-throughput computational techniques successfully predicted many new ternary nitrides,\cite{sun_map_2019} and combinatorial co-sputtering has proven to be a powerful tool for experimentally realizing these predicted materials.\cite{zakutayev2022experimentalSynthesisNitrides, zhuk2021synthesisZn2VN3, heinselman_thin_2019, greenaway2020combinatorialMgSnN2, greenaway2022zincPhotoelectrochemicalZnTiN2, bauers2019ternaryrocksaltsemiconductors, sun_map_2019, woods2022roleZnZrN2, smaha2022synthesisAlGdN, talley2021synthesisLaWN3, yu2023balanceHfWN, zakutayev2021synthesisZn2NbN3, kikuchi2021theoreticalYZn3N3}

In particular, Mg and Zn can be combined with early transition metals and main-group elements to form ternary nitrides that are structurally-related to the binary \ce{$M$^{3+}N} nitrides (e.g., wurtzite GaN or rocksalt TiN).\cite{zakutayev2022experimentalSynthesisNitrides} 
With $A$ as Zn or Mg and $M$ as a main group or transition metal, the stoichiometries \ce{$A$^{2+}$M$^{4+}N2}, \ce{$A$^{2+}_2$M$^{5+}N3}, and \ce{$A$^{2+}_3$M$^{6+}N4} have the 1:1 cation:anion ratio of \ce{$M$^{3+}N} compounds, and semiconducting properties emerge when $M$ is a $d^0$ transition metal or a main-group element.\cite{martinez2017synthesis_II_IV_V2, zakutayev2022experimentalSynthesisNitrides}
Furthermore, Zn and Mg favor 4- or 6-fold coordination, respectively, and therefore tend to produce the respective wurtzite-derived or rocksalt-derived structures. 
For example, exploration of the Zn-Mo-N phase space by combinatorial sputtering revealed a wurtzite-like structure across a range of compositions, from metallic \ce{ZnMoN2} to semiconducting wurtzite-like \ce{Zn3MoN4} (with a bandgap of 2.4~eV).\cite{arca2018redoxZnMoN2}
Similarly, the Mg-W-N phase space is a promising area of exploration because W has multiple possible oxidation states (between 0 and 6+, inclusive), potentially leading to varied structures and properties.

Combinatorial co-sputtering is a good choice to rapidly survey this potentially complex phase space.
However, materials discovered by combinatorial sputtering often deviate from those predicted by computational methods or synthesized in bulk on a key detail: cation (dis)order.\cite{zakutayev2022experimentalSynthesisNitrides, schnepf_utilizing_2020}
This discrepancy can potentially be beneficial, such as when cation-disorder lowers the bandgap into the visible range.\cite{schnepf_utilizing_2020, greenaway2022zincPhotoelectrochemicalZnTiN2}
In other cases, cation disorder negatively impacts optoelectronic properties by localizing charge carriers\cite{schnepf_utilizing_2020, lany2017monteCarloZnSnN2} or even leading to polymorphism.\cite{woods2022roleZnZrN2}
How to control this structural polymorphism and cation disorder is still an open question. For example, annealing conditions are known to affect the degree of cation (dis)order,\cite{blanton2017characterizationOrderZnGeN2, fetzer2002effectSb_GaInP_ordering} but that control is often material-specific and difficult to explore in a high-throughput manner.\cite{schnepf_utilizing_2020}
Therefore, understanding metastable phase formation and cation (dis)order in ternary nitrides remains a pressing challenge for the field to fully realize the tunable properties of this promising class of materials.

In this report, we describe the discovery of several new Mg-W-N compounds in this previously-unexplored ternary phase space.
We show that thin film combinatorial co-sputtering methods yielded cation-disordered rocksalt (RS, space group $Fm\bar{3}m$, $0.1<$ Mg/(Mg+W)$<0.9$) and hexagonal boron nitride structures (h-BN, space group $P6_3/mmc$, $0.7<$ Mg/(Mg+W)$<0.9$) covering the \ce{MgWN2} and \ce{Mg3WN4} stoichiometries.
In contrast, our bulk ceramic methods yielded cation-ordered \ce{MgWN2} with space group $P6_3/mmc$.  
We call this cation-ordered structure ``rocksaline'' (RL) as a portmanteau of the rocksalt-like Mg and nickeline-like W layers.
Thermodynamic calculations confirm that the RL polymorph is the ground state of the \ce{MgWN2} composition.
Rapid thermal annealing (RTA) of thin films converted the disordered RS structure to this ordered RL structure, in a narrow composition window near the \ce{MgWN2} stoichiometry, resolving this difference between the thin film and the bulk synthesis results.
The \ce{Mg3WN4} was only produced in thin films, and formed in either the RS or h-BN structure.
Thermodynamic calculations reveal these polymorphs to be close in energy to one another and slightly metastable.
Electronic structure calculations suggest that RL \ce{MgWN2} should be a semiconductor, while RS \ce{MgWN2} should be metallic.
Resistivity measurements of the synthesized films as a function of composition and temperature show both RL \ce{MgWN2} and RS \ce{Mg3WN4} are semiconducting, but were unable to verify the charge transport behavior of RS \ce{MgWN2}.
These findings show how RTA treatment of disordered films can build upon existing combinatorial co-sputtering techniques to rapidly assess the thermodynamic synthesizability of a predicted cation-ordered phase.

\section{Methods}
\subsection{Bulk structural measurements and analysis}
Powder X-ray diffraction (PXRD) measurements were performed using a Bruker DaVinci diffractometer with Cu K$\alpha$ X-ray radiation. All samples were prepared for PXRD from within the glovebox by placing powder on off-axis cut silicon single crystal wafers to reduce the background, and then covered with polyimide tape to slow exposure to the atmosphere.
However, as PXRD showed that the product (\ce{MgWN2}) is air stable, a PXRD pattern was collected without tape to minimize the large scattering background (Figure \ref{fig:bulk_MgWN2}). 

Full-pattern fitting of thin film XRD, GIWAXS, and PXRD data was performed using TOPAS v6.\cite{coelho2018topas}
For thin film samples, 2D diffraction images showed texturing (i.e., preferred orientation), meaning that integrated peak intensities may not directly correspond to electron density. 
Therefore, we performed LeBail fits using the appropriate space group and refined lattice parameters and crystallite size broadening. 
For the \ce{MgWN2} phase in the RL structure, a model was created by substituting W for Mo in the previously reported \ce{MgMoN2} structure in space group $P6_3/mmc$.\cite{verrelli_viability_2017} Rietveld analysis was then performed to refine the lattice parameters, crystallite size broadening, and site occupancy. 
In all cases, 10-term polynomial functions were refined to fit the background.
Structural visualizations were performed with VESTA.\cite{momma2011vesta}

\subsection{Thin film synthesis and annealing experiments}
Combinatorial co-sputtering of Mg-W-N film libraries were conducted in two custom vacuum chambers, both with base pressures of $<10^{-7}$~Torr. 
Mg and W targets (2~inch diameter, Kurt J. Lesker, 99.95\% purity) were angled towards a stationary substrate and sputtered using radiofrequency (RF) excited plasma of the Ar/\ce{N2} gas mixture in the chamber. 
Sputter powers ranged from 30~W to 90~W for each target, to shift the Mg/(Mg+W) ratio across the whole composition window.
Gases were introduced at 50~sccm Ar and 50~sccm \ce{N2}, with a 10~Torr process pressure during deposition. 
The N plasma intensity was enhanced by RF plasma source at 350~W. 
Most samples were deposited on 2~inch by 2~inch (001)-oriented Si substrates. 
Select samples were deposited on insulating substrates (e.g., 100~nm \ce{SiO2} on Si or 100~nm \ce{SiN$_x$} on Si) for electronic property measurements, as indicated in the text. 
Select samples were coated with a 15~nm TiN capping layer, sputtered from a 2~inch diameter Ti target, to protect against atmospheric exposure.
During these capping depositions, the substrate was rotated to ensure a homogeneous capping layer.
A diagram for this experimental setup is shown in Figure \ref{fig:experimental_setup}A.

Rapid thermal annealing (RTA) experiments were conducted on individual compositionally-graded library rows in flowing \ce{N2} atmosphere at ambient pressure. Heating profiles started with a +100~\textdegree{}C/min ramp to 100~\textdegree{}C and a 3~min dwell to drive off adsorbed water, followed by a +100~\textdegree{}C/min ramp to a $T_\mathrm{anneal}$ set-point in the 600-1200~\textdegree{}C range for a 3~min dwell. Samples were cooled by turning off the heating source. A diagram for this experimental setup is shown in Figure \ref{fig:experimental_setup}B. 

\subsection{Thin film composition and structure}
Combinatorial libraries were measured using the standard 4$\times$11 grid employed at NREL, with data analysis conducted using the COMBIgor software package.\cite{talley2019combigor} Each library was mapped with X-ray diffraction (XRD) using a Bruker D8 Discover with Cu K$\alpha$ radiation and an area detector. 
Select samples were measured by high-resolution synchrotron grazing incidence wide angle X-ray scattering (GIWAXS) at the Stanford Synchrotron Radiation Lightsource (SSRL) at a wavelength of 0.9744~\AA{} with a Rayonix MX225 CCD area detector, a 3\textdegree{} incident angle, and a 50 $\mu$m  $\times$ 150 $\mu$m spot size.  
GIWAXS detector images were integrated with GSAS-II.\cite{toby2013gsas} 

Compositional analysis was performed with X-ray fluorescence (XRF) and Rutherford Back-Scattering (RBS). 
Metal ratios were mapped using a Bruker M4 Tornado XRF with a Rh source operating at 50~kV and 200~$\mu$A. 
The spot size was 25~$\mu$m in diameter. 
The measurements were performed under vacuum ($<$20~mbar) with an exposure time of 200~s for each measurement.  
Nitrogen and oxygen ratios for select samples were quantified with RBS. RBS was run in a 168\textdegree{} backscattering configuration using a model 3S-MR10 RBS system from National Electrostatics Corporation with a 2 MeV \ce{He+} beam energy. 
Samples were measured for a total integrated charge of 160 $\mu$C. 
RBS spectra were modeled with the RUMP software package.\cite{barradas2008summaryRBS}

\subsection{Thin film property measurements}
Room temperature resistivity was measured on thin films using a custom-built collinear four-point probe instrument by sweeping current between the outer two pins while measuring voltage between the inner pins (1~mm between each pin). Conventional geometric corrections were applied to convert the measured resistance into sheet resistance and then resistivity.\cite{smits1958measurement} The measured films were deposited on insulating substrates (either 100~nm thick \ce{SiO2} on Si or 100~nm thick \ce{SiN$_x$} on Si) to avoid contribution from the substrates. 

Temperature-dependent electrical resistivity was measured on thin films using a Lake Shore Cryotronics Model 8425. 
Small squares (5~mm $\times$ 5~mm) were cleaved out of libraries deposited on insulating substrates. 
For compositions near \ce{MgWN2}, indium contacts were pressed into the film near the corners of the squares. 
Indium contacts were non-ohmic on \ce{Mg3WN4} films, so Ti/Au contacts were deposited by evaporation. 
Temperature-dependent sheet resistance was measured from 104~K to 298~K for most samples, with RL \ce{MgWN2} measured from 36~K to 298~K. 
Resistivity was calculated using XRF-measured film thickness.

\subsection{Bulk synthesis}
Powders of \ce{Mg3N2} (Alfa Aesar, $>99.6$\%, 325 mesh) and \ce{W} (Sigma-Aldrich, 99\%, 42~$\mu$m) were used as received. 
As these reagents are air sensitive, they were prepared and stored in an argon-filled glovebox (\ce{O2} $<$ 0.1 ppm, \ce{H2O} $<$ 0.5 ppm). 
Bulk reactions were prepared by grinding together the reagent powders with an agate mortar and pestle, pelletizing the mixture by cold-pressing in a 0.25~in die at 300~MPa (approximately 100--200~mg per pellet), loading the pellet into a cylindrical alumina crucible held horizontally in an alumina boat, and loading the boat into a mullite or quartz process tube. 
A Zr foil cap was fit into the mouth of the alumina crucible to decrease \ce{Mg3N2} loss by volatization and to sacrificially react with any trace \ce{O2}. 
Without air exposure, the samples were reacted in a tube furnace under flowing purified \ce{N2} (ca. 20~mL/min flow rate). 
A diagram for this system is shown in Figure \ref{fig:experimental_setup}C.
Reactions were conducted by heating the sample at +10~\textdegree{}C/min to the dwell temperature, dwelling for approximately 5--20~h at various temperatures up to 1100~\textdegree{}C, and then cooling by switching off the furnace. 
Samples were recovered into the Ar glovebox. 
This procedure was adapted from the strategy used by Verrelli,~et al.,~to synthesize \ce{MgMoN2}.\cite{verrelli_viability_2017}

\subsection{Computational methods}
Formation energies were calculated using density functional theory (DFT) using the corrected generalized gradient approximation (GGA+U) implemented in the Vienna Ab initio Structural Package (VASP).
These calculated values were sourced from the Materials Project when available (v2021.11.10).\cite{jain2011formationEnergies, jain2013commentaryMaterialsProject} 
Calculations for additional structures that were not already in the Materials Project database (i.e., all \ce{MgWN2} polymorphs, RS and h-BN \ce{Mg3WN4}) were conducted using Atomate (v1.0.3)\cite{mathew2017atomate} and Fireworks (v2.0.2)\cite{jain2015fireworks} to execute the structure optimization workflow with compatibility with Materials Project entries.
Calculations were carried out on cation-ordered versions of the experimentally observed cation-disordered structures. 
Pymatgen (v2022.4.19) was used to construct the ternary phase diagram shown in Figure \ref{fig:MgWN_Hull}.\cite{ong2013pymatgen}

\section{Results and Discussion} 

\subsection{Bulk synthesis of cation-ordered \ce{MgWN2}}
\begin{figure}
    \centering
    \includegraphics[width = 3.2 in]{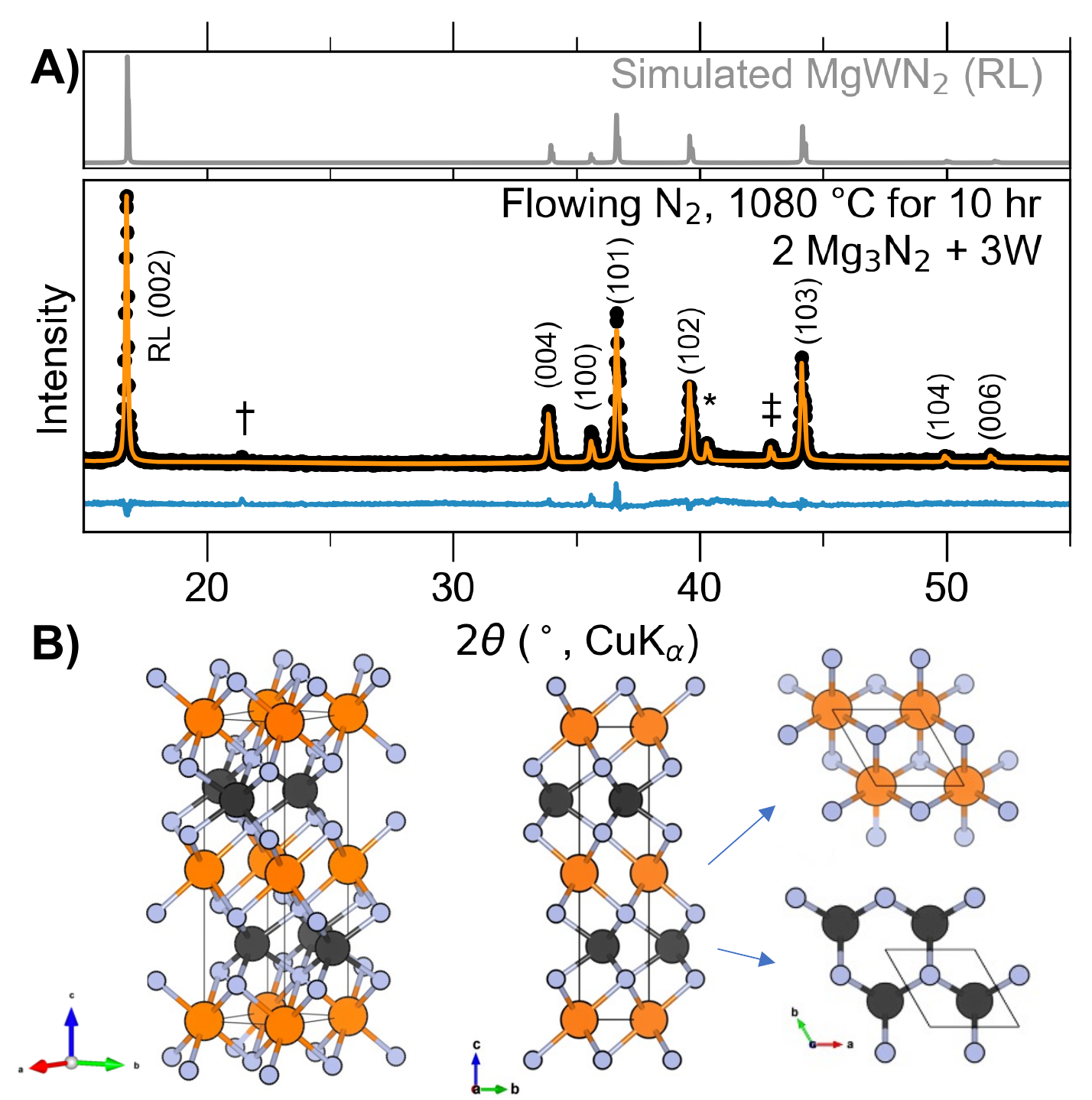}
    \caption{A) Rietveld refinement (orange trace) of a PXRD pattern (black dots) of \ce{MgWN2} produced by heating \ce{2 Mg3N2 + 3 W} (100\% excess \ce{Mg3N2}) under flowing \ce{N2} for 10~h at 1080~\textdegree{}C. The difference trace is shown in blue. The simulated pattern of RL \ce{MgWN2} is shown for reference in the box above. W (*), MgO (‡), and an unidentified phase ($\dagger{}$) are trace impurities in the pattern. 
    B) Ball-and-stick models of RL \ce{MgWN2} from different perspectives.}
    \label{fig:bulk_MgWN2}
\end{figure}

Bulk syntheses yielded \ce{MgWN2} in a cation-ordered layered hexagonal crystal structure (Figure \ref{fig:bulk_MgWN2}) previously reported for \ce{MgMoN2}.\cite{verrelli_viability_2017, wang2012solidMgMoN2} 
We call this structure ``rocksaline'' (RL) for short, a portmanteau of rocksalt and nickeline, because this structure has interleaved layers of octahedrally-coordinated \ce{Mg^{2+}} (rocksalt-like) and \ce{W^{4+}} in a trigonal-prismatic site (nickeline-like). 
The RL \ce{MgWN2} phase formed as a black powder from a reaction between \ce{Mg3N2} and W powders in a 2:3 ratio heated at 1080~\textdegree{}C for 10~h. As the balanced reaction is \ce{Mg3N2 + 3 W + 2 N2 -> 3 MgWN2}, this synthesis requires a full excess equivalent of \ce{Mg3N2} to proceed to completion. Still, W often persisted as an impurity owing to the volatility of Mg at elevated temperatures and the refractory nature of W.
Syntheses conducted at lower temperatures did not induce reaction, suggesting a significant kinetic barrier to reactivity between \ce{Mg3N2} and W. 

Crystallographic analysis via refining the degree of site inversion ($x$) for (\ce{Mg_{1-$x$}W_{$x$})(W_{1-$x$}Mg_{$x$})N2} using the Rietveld method leads to $x$ = 0.115(10), suggesting some cation disorder (Table \ref{tab:MgWN2_RL_xtal}). 
For comparison, $x$ = 0.5 would indicate complete cation disorder, and $x$ = 0 would indicate a fully ordered phase.
However, site occupancy is modeled by fitting relative peak intensities, and peak intensities also vary with preferred orientation which may be present in these data but which were not included in the model.\cite{dinnebier2018rietveld} 
Cation ordering is most clearly defined by a (002) reflection at $2\theta = 17$\textdegree{} (Figure \ref{fig:simulated_RL_ordering}), and the strong reflection observed in Figure \ref{fig:bulk_MgWN2} suggests a substantial degree of cation ordering.
The isostructural \ce{MgMoN2} synthesized by the same method was modeled to be fully ordered by combined analysis of synchrotron PXRD and neutron powder diffraction data.\cite{verrelli_viability_2017}

The formation of RL \ce{MgWN2} by high-temperature ceramic synthesis indicates that the RL polymorph defines the thermodynamic ground state. 
Excess \ce{Mg3N2} used in bulk syntheses did not lead to any signs of a more Mg-rich phase (i.e., \ce{Mg3WN4}), so we hypothesize any ordered configurations of those materials (e.g., an ordered wurtzite structure, $Pmn2_1$) may be destabilized at the elevated temperatures (and thus lower nitrogen chemical potential, $\mu_\mathrm{N}$) required for ceramic synthesis.
The bulk synthesis results differed from the the thin-film work presented next, showing the contrast between different precursor options: diffusion-limited bulk-powders compared to atomically-dispersed films. 

\subsection{Synthesis of Mg-W-N thin films by combinatorial co-sputtering}
\begin{figure}
    \centering
    \includegraphics[width = 3.2 in]{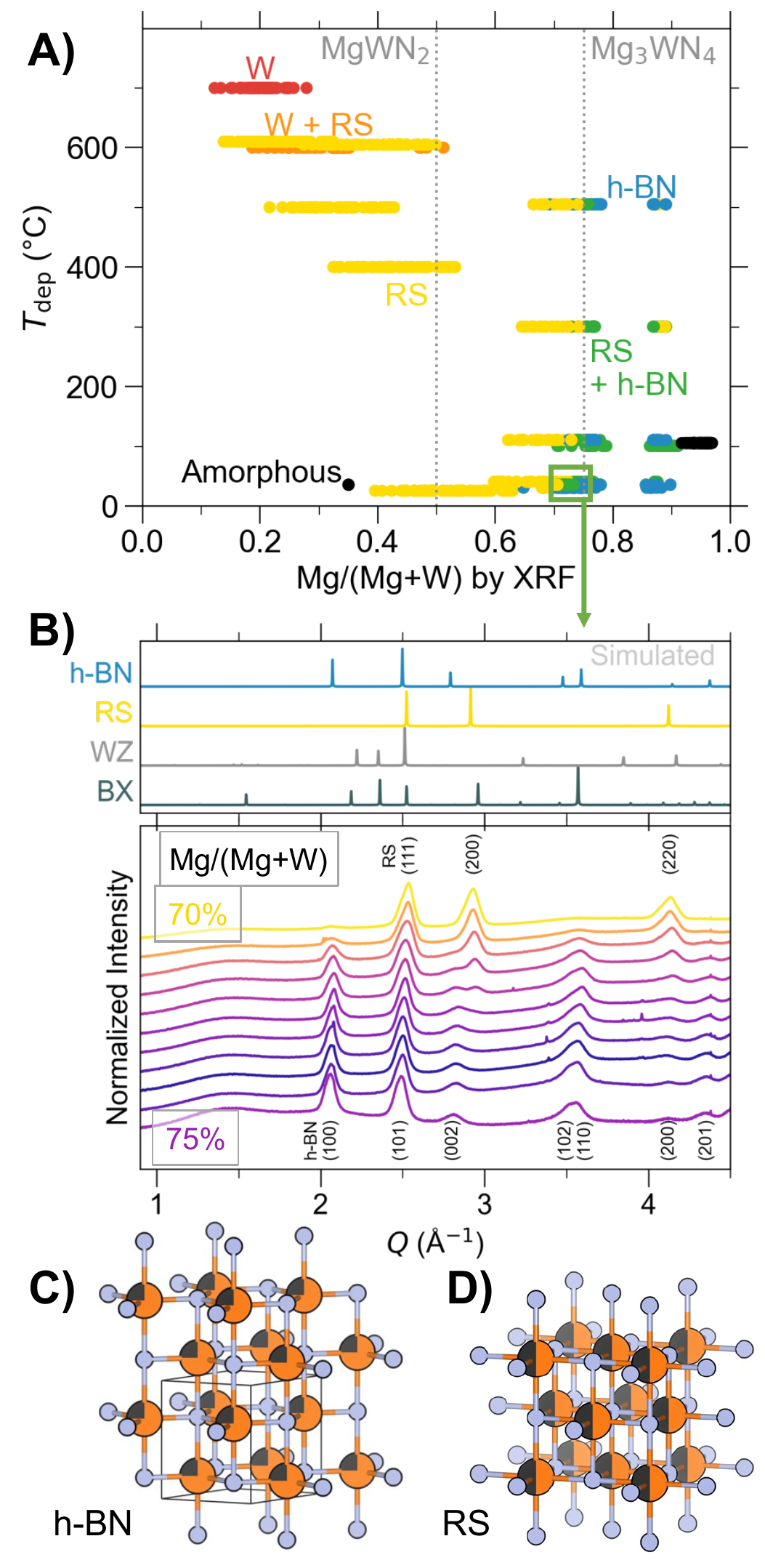}
    \caption[PXRD results of Mg-W-N library depositions]{A) Phase diagram of thin film Mg-W-N extracted from combinatorial growths at various temperatures ($T_\mathrm{dep}$).
    B) GIWAXS patterns from a library deposited at ambient conditions, showing the transition between the rocksalt (RS) and h-BN structures. The wurtzite (WZ) and anti-bixbyite (BX) structures are not observed.
    Ball-and-stick models of C) h-BN \ce{Mg3WN4}. D) RS \ce{MgWN2}.
    }
    \label{fig:combi_MgWN}
\end{figure}
  
Combinatorial co-sputtering from Mg and W targets in a \ce{N2}/Ar environment resulted in cation-disordered phases with either the RS or the h-BN structure, as determined by laboratory XRD (Figure \ref{fig:combi_MgWN}). 
The RS structure shows the greatest degree of stability, crystallizing across a wide range of compositions ($0.1<$ Mg/(Mg+W)$<0.9$) and substrate temperatures (up to 600~\textdegree{}C). 
At elevated substrate temperatures (ca. 700~\textdegree{}C), Mg volatilizes, leaving behind metallic W. 
At Mg/(Mg+W) ratios near 0.75 (i.e., \ce{Mg3WN4}), a h-BN structure is observed in some libraries; it was characterized in greater detail by GIWAXS (Figure \ref{fig:combi_MgWN}B).
This h-BN structure only appeared in depositions using one of the custom vacuum chambers, but not the other. This suggests a subtle (and yet-undetermined) process parameter, such as nitrogen-plasma density or oxygen content, may play a role.
Even within the one chamber that yielded h-BN \ce{Mg3WN4}, some Mg-rich samples still show the RS structure, suggesting these two polymorphs may be close in energy.
Other Mg-rich points did not exhibit any crystalline phases and are marked as amorphous in Figure \ref{fig:combi_MgWN}A.

The coexistence of h-BN and RS polymorphs near the \ce{Mg3WN4} stoichiometry suggests the phases may be energetically similar for this Mg/(Mg+W) ratio. 
Indeed, they are structurally related, with the h-BN structure being an intermediate in a displacive transformation between the RS and WZ structures.\cite{limpijumnong2001theoreticalRS_to_BN_to_WZ} 
This h-BN structure is uncommon among ternary nitrides. 
The only prior report we can identify in literature is that of Zn-rich compositions for \ce{ZnZrN2}.\cite{woods2022roleZnZrN2} 
However, the five-fold coordination environment of the h-BN is analogous to the transition state experienced by WZ-type ferroelectric materials (e.g., Al$_{1-x}$Sc$_x$N) as they undergo switching.\cite{fichtner2019_ferroelectricAlScN}
As another example of a similar motif, \ce{Mg3Al3N5} has an \ce{Al^{3+}} ion split across two face-sharing tetrahedral sites,\cite{schmidt2017crystalMg3AlN3} which is structurally similar to the WZ $\rightarrow$ h-BN $\rightarrow$ WZ displacement of ferrolectrics.
Lastly, a prior study predicted the ground state for \ce{Mg2NbN3} and \ce{Mg2TaN3} to be this h-BN structure type,\cite{zakutayev2022experimentalSynthesisNitrides} although sputtering experiments subsequently showed that \ce{Mg2NbN3} crystallizes as a cation-disordered rocksalt.\cite{bauers2019ternaryrocksaltsemiconductors, todd2021twostep}
The infrequent occurrence of this polymorph suggests decreased stability relative to other high-symmetry phases like the RS polymorph, a hypothesis supported by our RTA experiments (Figure \ref{fgr:MgWN_RTA_hBN_to_RS}) and inability to produce it in bulk.

\subsection{Rapid thermal annealing of combinatorial libraries}

\begin{figure}
    \centering
    \includegraphics[width = 3.2 in]{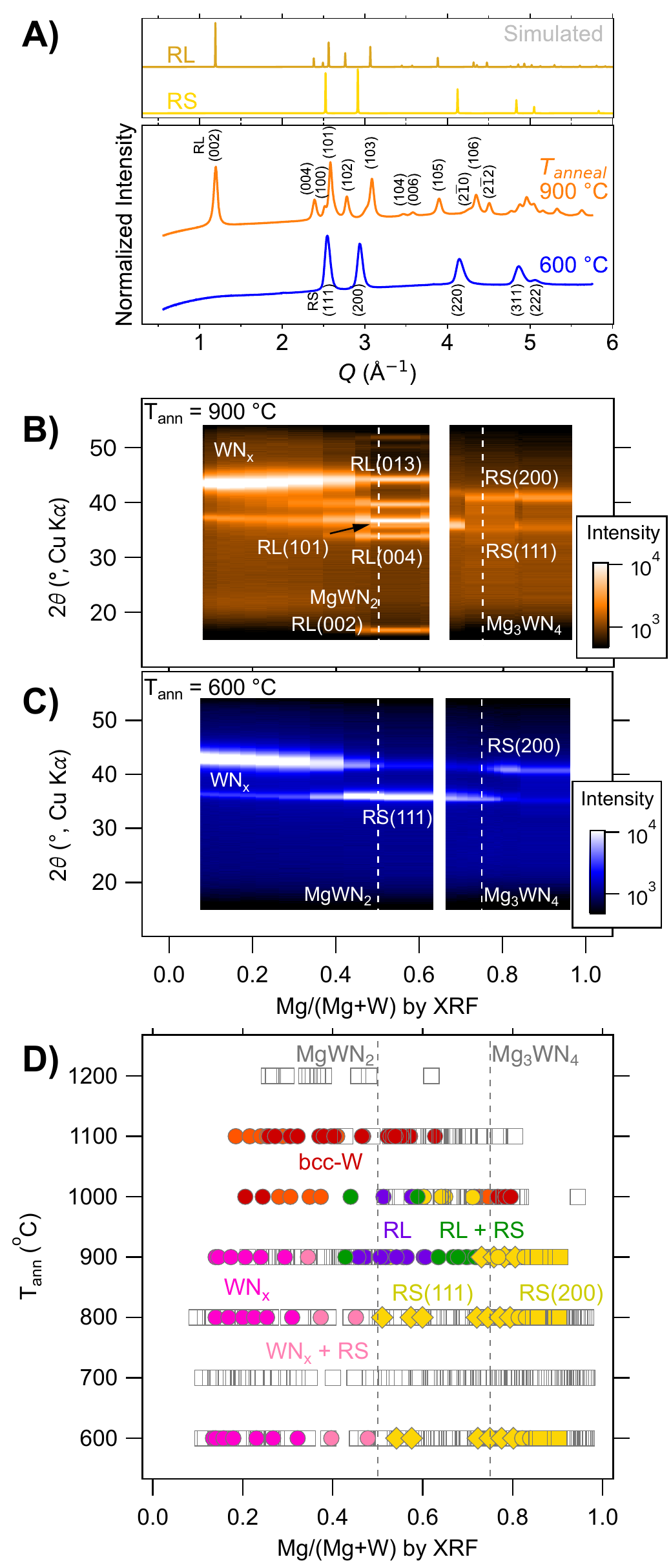}
    \caption{
    A) Synchrotron GIWAXS patterns of \ce{MgWN2} annealed at 600~\textdegree{}C and 900~\textdegree{}C. 
    B, C) Laboratory XRD heatmaps as a function of Mg/(Mg+W) for library rows annealed at 600~\textdegree{}C and 900~\textdegree{}C. Labels for the (100) and (102) RL \ce{MgWN2} reflections are omitted for clarity.
    D) Phase diagram of Mg-W-N depositions as a function of annealing temperature. Samples which were manually identified are indicated by colored markers. Empty markers were not manually labeled but were measured by XRF and XRD, and phases can be inferred from neighboring points.
}
    \label{fig:MgWN_rapidThermalAnnealing}
\end{figure}

RTA experiments of combinatorial film libraries show that annealing can induce cation ordering near the \ce{MgWN2} stoichiometry (Figure \ref{fig:MgWN_rapidThermalAnnealing}). 
The samples near the stoichiometric \ce{MgWN2} composition retained the RS structure at $T_\mathrm{anneal}$ = 600~\textdegree{}C, but a clear structure transition to the RL polymorph occurred by $T_\mathrm{anneal}$ = 900~\textdegree{}C (Figure \ref{fig:MgWN_rapidThermalAnnealing}A). This indicates that the as-deposited RS structure is kinetically-stable up to moderately high temperatures (ca. 600~\textdegree{}C). 
High temperatures (ca. 900~\textdegree{}C) are needed to allow local diffusion of the randomly-dispersed metals in octahedral environments (the RS structure) to their energetically-preferred coordination environments (octahedral \ce{Mg^{2+}} and trigonal-prismatic \ce{W^{4+}} in the RL structure. 

For Mg-poor compositions (Mg/[Mg+W]$<0.4$), annealing produces a slightly different structure than the RS observed in depositions at elevated temperatures, a structure we call \ce{WN$_x$}.
XRD patterns show two reflections that are similar to the RS (111) and (200) reflections, but which are spaced by slightly too large a gap in $2\theta$ to be consistent with the $Fm\bar{3}m$ structure (Figure \ref{fig:Mg_poor_phases}).
However, we are not able to precisely identify the space group of this phase. 
Only two reflections were detected, and diffraction images show substantial texturing, which suggests that additional reflections may exist outside the measured $\chi$ range.
Furthermore, the W-N binary system is complex, with 13 unique structures reported in the Inorganic Crystal Structure Database (ICSD) ranging in composition from \ce{W2N} to \ce{WN2}.
\cite{feng2019role_WN_Fm-3m, wang2017synthesis_delta_WN, sasaki2019highPressure_WN, wang2012synthesis_WN, chang2021crystal_W7N12}  
Given this complexity and ambiguity, we simply refer to these Mg-poor phases as \ce{WN$_x$}. 
This difference may stem from the elevated nitrogen chemical potential present in combinatorial depositions but absent during annealing, which may affect how much nitrogen is present in the film.\cite{caskey2014thinCu3N, zakutayev2021synthesisZn2NbN3} 
However, annealed samples labeled RS in Figure \ref{fig:MgWN_rapidThermalAnnealing}D (i.e., those with Mg/[Mg+W] $\geq$ 0.5) are well fit with the $Fm\bar{3}m$ space group.

The RS to RL transformation only occurs in a narrow composition window near Mg/(Mg+W) = 0.5 (i.e., \ce{MgWN2}, Figure \ref{fig:MgWN_rapidThermalAnnealing}D). For Mg-poor compositions with Mg/(Mg+W)$ < 0.42$ and Mg-rich compositions with Mg/(Mg+W)$ > 0.62$, the \ce{WN$_x$} and RS structures persisted at $T_\mathrm{anneal}$ = 900~\textdegree{}C. This shows that the ordered RL structure has a narrow compositional tolerance, while the \ce{WN$_x$} and RS structures can accommodate a large degree of off-stoichiometry. These results, along with the thermodynamic calculations presented next (Figure \ref{fig:MgWN_Hull}) confirm that the RL phase is the thermodynamic ground state up to approximately 1000~\textdegree{}C, as initially shown by bulk syntheses. 

\subsection{Thermodynamic analysis}
\begin{figure}
    \centering
    \includegraphics[width = 3.25in]{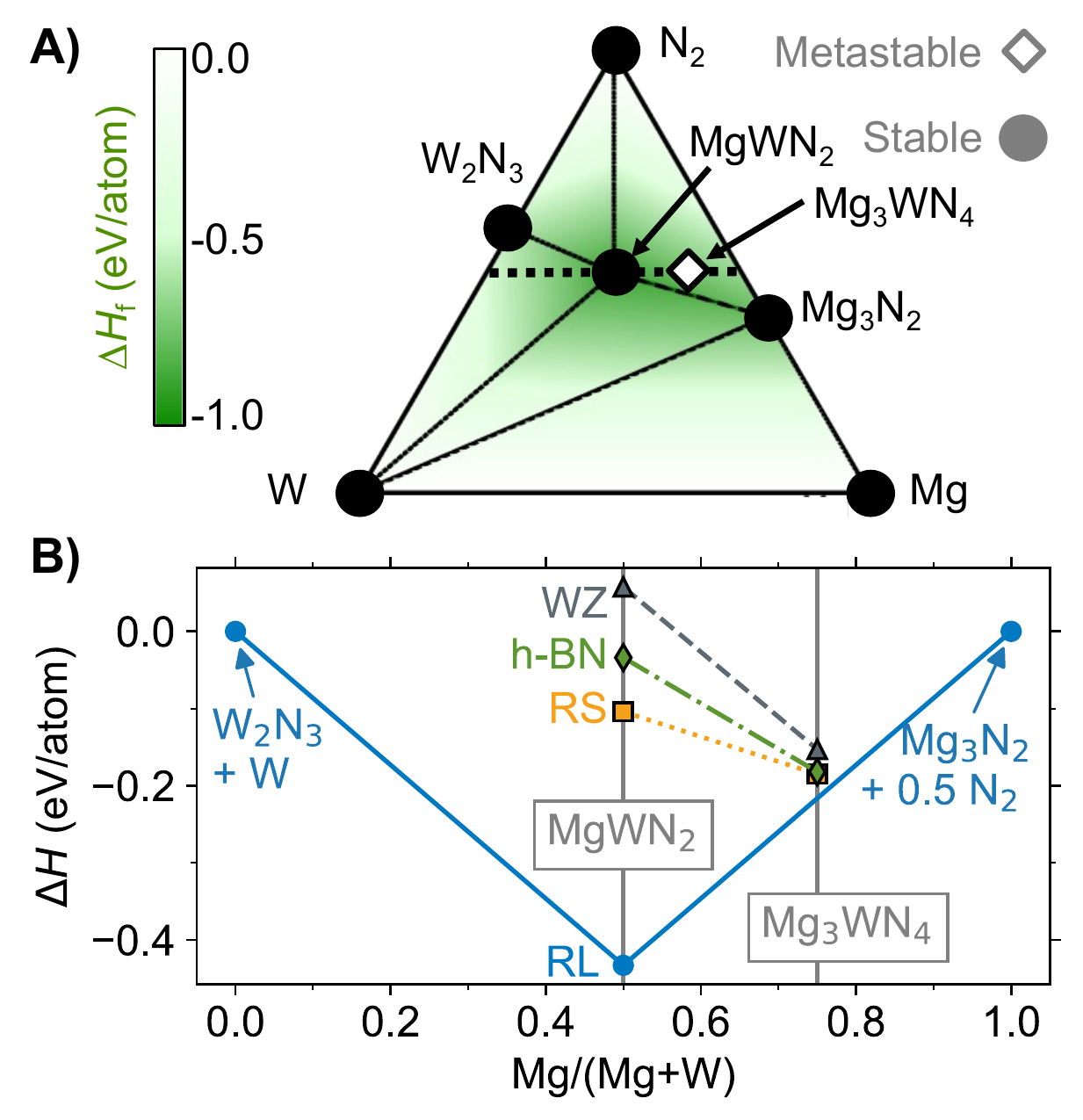}
    \caption[Calculated ternary and pseudobinary phase diagrams for the Mg-W-N system.]{A) Ternary phase diagram for the Mg-W-N system calculated using pymatgen.\cite{ong2013pymatgen} B) Pseudobinary isopleth calculated with a 1:1::(Mg+W):N ratio (corresponding to the black dotted trace from A). The vertical axis shows the relative formation energy ($\Delta H$) at $T = 0$~K compared to the most stable point in the binary hulls at this cation:anion ratio (\ce{W2N3 + W} and \ce{Mg3N2 + 1/2N2}). 
    Several highly metastable ternary phases in the NREL matDB and Materials Project databases are omitted for clarity.\cite{stevanovic2012NRELmatDB, jain2013commentaryMaterialsProject} 
    }
    \label{fig:MgWN_Hull}
\end{figure}

Calculated formation energies relative to the binaries show that RL \ce{MgWN2} is the only thermodynamically stable ternary in the Mg-W-N system, according to DFT calculations of the cation-ordered structures (Figure \ref{fig:MgWN_Hull}). 
The striking favorability of the RL polymorph of \ce{MgWN2} is driven by the electronic preference of $d^2$ metals (like \ce{W^{4+}}) for trigonal-prismatic coordination environments.\cite{kertesz1984octahedral}
The next lowest energy polymorph for \ce{MgWN2} is RS, followed by h-BN, then WZ. 
In the case of the \ce{Mg3WN4} stoichiometry, all three polymorphs (RS, h-BN, and WZ) are much closer to the hull than the metastable \ce{MgWN2} polymorphs. 
A RS \ce{Mg3WN4} structure (space group $I4/mmm$) is closest to the hull (+0.031~eV/atom above the hull), but the h-BN structure is only slightly higher in energy (+0.034~eV/atom above the hull).
The WZ-derived phase of \ce{Mg3WN4}, with a desirable predicted bandgap of ca.~5~eV,\cite{greenaway2021ternaryReview} is only slightly higher (+0.063~eV/atom above the hull).

The DFT calculations shown in Figure \ref{fig:MgWN_Hull} agree with our synthetic results. 
RL \ce{MgWN2} was the only ternary phase formed by bulk synthesis, where high temperatures are sufficient to overcome kinetic barriers to produce thermodynamic ground-state phases. 
The formation of RS \ce{MgWN2} by combinatorial sputtering is also consistent with the trend from calculations and with prior literature.\cite{woods2022roleZnZrN2, bauers2019ternaryrocksaltsemiconductors} 
In the case of physical vapor deposition methods (like sputtering), atoms arrive at the film surface in a disordered configuration (i.e., high effective temperature). 
Under these conditions, configurational entropy favors structures with a single type of cation site (like RS, h-BN, and WZ) and enthalpy penalizes structures with two or more distinct cation sites (like RL), as demonstrated for the Zn-Zr-N system.\cite{woods2022roleZnZrN2}
In other words, RS is a disorder-tolerant structure that becomes energetically favorable under sputtering synthesis conditions. 
While we do not consider disorder in the calculations shown in Figure \ref{fig:MgWN_Hull}B, the ordered RS phase is lower in energy than the ordered WZ or h-BN phases, suggesting \ce{Mg^{2+}} and \ce{W^{4+}} prefer octahedral coordination environments over tetrahedral (WZ) and trigonal bipyramidal (h-BN) environments.
Lastly, oxygen substitution on nitrogen sites is common in nitrides,\cite{greenaway2021ternaryReview, schnepf_utilizing_2020} and these materials are no exception. 
RBS measurements detect O/(N+O) = 15\% for \ce{Mg3WN4} with a h-BN structure (Figure \ref{fig:RBS_ratios}). 
Auger electron spectroscopy measurements on \ce{Mg3WN4} with a RS structure detect lower levels of oxygen (O/(N+O)$<2$\%, Figure \ref{fig:aes}).
These measurements suggest that oxygen incorporation may stabilize the h-BN structure over the RS structure for \ce{Mg3WN4}. 
Oxygen impurities affect the energy landscape but are not accounted for in these calculations.

\subsection{Electronic properties}
\begin{figure}
    \centering
    \includegraphics[width = 3.25 in]{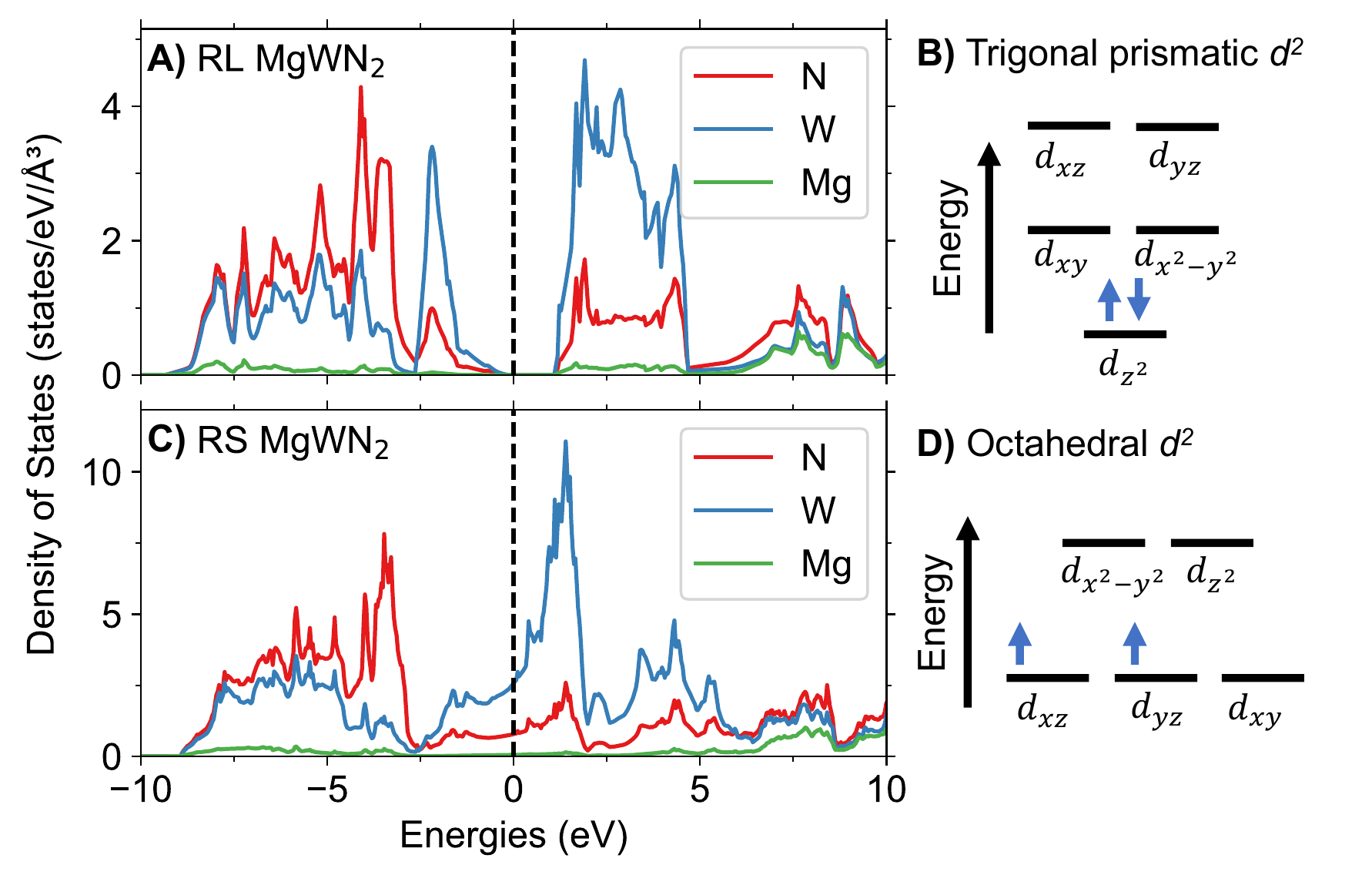}
    \caption{Calculated density of states (DoS) for the A) RL \ce{MgWN2} and C) RS \ce{MgWN2} (calculated using the ordered structure with space group $I4_1/amd$). Ligand field splitting diagrams for \ce{W^{4+}} in B) trigonal prismatic and D) octahedral environments.}
    \label{fig:dos_RS_v_RL}
\end{figure}

The polymorphic differences for \ce{MgWN2} should lead to different properties.
To assess this possibility, we conducted electronic structure calculations on the cation-ordered RL polymorph and a cation-ordered model of RS \ce{MgWN2}.
As these electronic structure calculations cannot be conducted on disordered models, we created a cation-ordered RS \ce{MgWN2} phase based on the \ce{$\gamma$-LiFeO2} structure type (space group $I4_1/amd$).
Calculated density of states (DoS) diagrams show that RS \ce{MgWN2} has states at the Fermi level and should exhibit metallic behavior, while RL \ce{MgWN2} is calculated to be a semiconductor with a 1.18~eV bandgap (Figure \ref{fig:dos_RS_v_RL}).
This latter finding is consistent with the 0.7~eV bandgap calculated for RL \ce{MgMoN2} (albeit that phase was calculated without the use of hybrid functionals),\cite{verrelli_viability_2017} and with the band structure of \ce{MoS2}, where \ce{Mo^{4+}} takes a trigonal-prismatic coordination environment.\cite{kasowski1973bandMoS2, mattheiss1973energyMoS2}
Band structure diagrams are shown in Figures \ref{fig:bandsMgWN2_RL} and \ref{fig:bandsMgWN2_RS}. 

This difference can be rationalized via a simple ligand field splitting model. 
The RL polymorph has the 5$d^2$ valence electrons fully occupying a $d_{z^2}$ orbital (Figure \ref{fig:dos_RS_v_RL}B).
The lowest unoccupied orbitals are degenerate $d_{x^2-y^2}$ and $d_{xy}$, suggesting a bandgap defined by $d-d$ transitions. 
In contrast, the \ce{W^{4+}} in the RS polymorph undergoes octahedral ligand field splitting. That leads to metallic conductivity via three degenerate orbitals ($d_{xy}$, $d_{xz}$, and $d_{yz}$) for the 5$d^2$ valence electrons (Figure \ref{fig:dos_RS_v_RL}D).
Such splitting is consistent with the calculated DoS, where W states make up a large fraction of the valence and conduction bands for RL \ce{MgWN2} and states near the Fermi level for RS \ce{MgWN2}.

\begin{figure}
    \centering
    \includegraphics[width = 3.25 in]{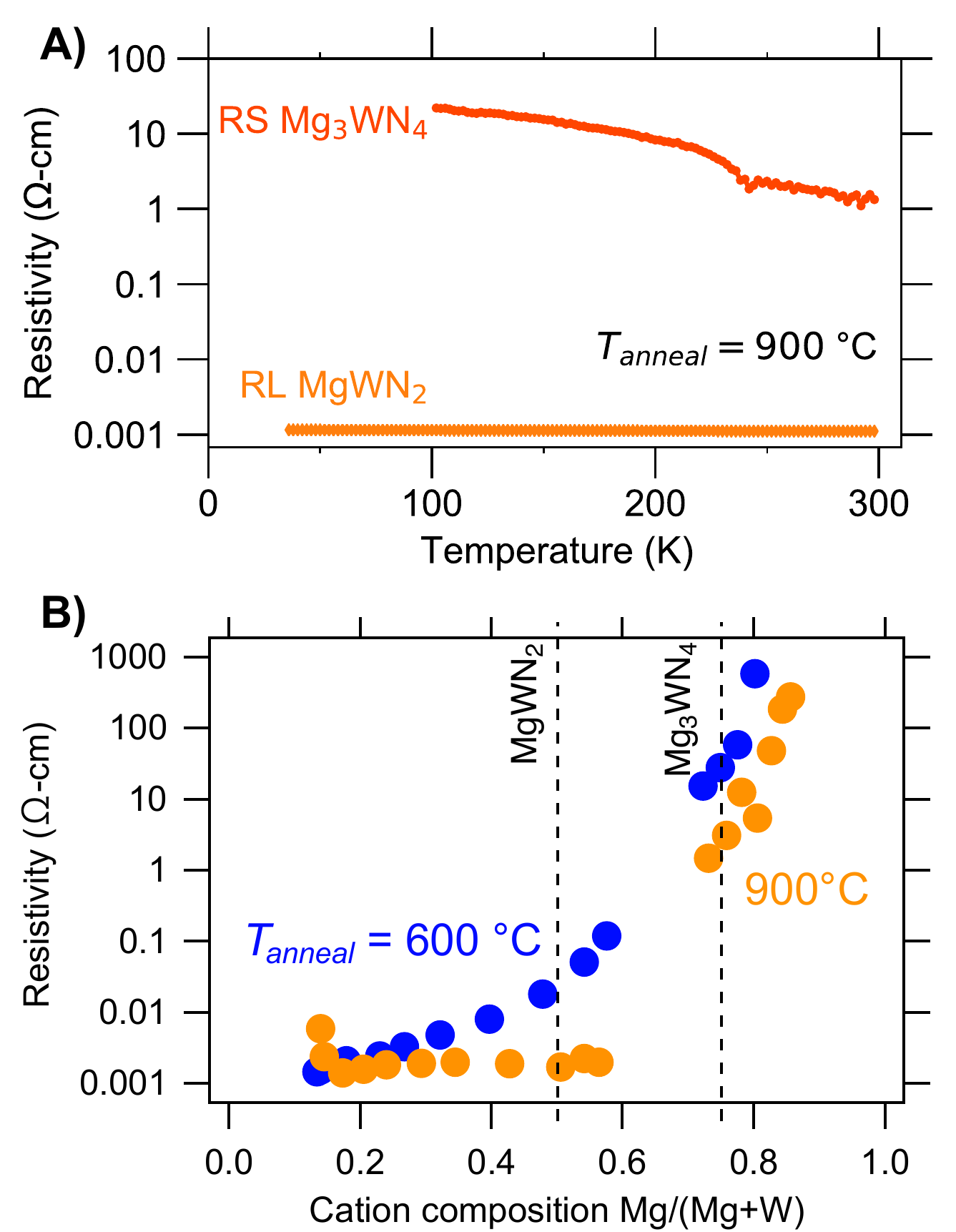}
    \caption[Electronic properties of Mg-W-N films]{
    A) Temperature-dependent resistivity measurements of select samples of RS \ce{Mg3WN4} and RL \ce{MgWN2} ($T_\mathrm{anneal} = 900$~\textdegree{}C).
    B) Colinear four-point probe measurements of Mg-W-N library rows annealed at 600~\textdegree{}C and 900~\textdegree{}C.
    }
    \label{fig:MgWN2_resistivity}
\end{figure}
Temperature-dependent resistivity measurements of thin films indicate semiconducting behavior for RL \ce{MgWN2} and RS \ce{Mg3WN4} (Figure \ref{fig:MgWN2_resistivity}A). 
Resistivity decreases with increasing temperature for both samples, although the trend for \ce{MgWN2} is significantly weaker than for \ce{Mg3WN4} (Figure \ref{fig:resistivity_MgWN2_RL}).
This trend suggests thermally activated charge transport.
The semiconductivity of RS \ce{Mg3WN4} is consistent with the 6+ oxidation state for W in that phase (5$d^0$ electron configuration).
The change in slope near 230 K is an artefact of the instrument.\cite{smaha2023structuralLaWN3}
The resistivity of RL \ce{MgWN2} is low (ca. $0.001~\Omega$-cm), suggesting a high level of doping and/or a small bandgap. 
The resistivity of RS \ce{Mg3WN4} is substantially larger, indicating a lower dopant content and/or a large bandgap.
We were not able to reliably measure temperature-dependent resistivity of RS \ce{MgWN2}, possibly owing to compositional gradients within the film or sample degradation from air exposure over time.
Similar trends in conductivity were observed in the Zn-Mo-N system, where films of a wurtzite structure spanned low-resistivity \ce{ZnMoN2} to insulating \ce{Zn3MoN4}.\cite{arca2018redoxZnMoN2} 

Electronic properties of these films are affected by film quality and composition. 
Room temperature resistivity measurements show that annealing at 900~\textdegree{}C decreases resistivity slightly across the whole composition range (compared to samples annealed at 600~\textdegree{}C), consistent with decreased grain-boundary resistance (Figure \ref{fig:MgWN2_resistivity}B).
Additionally, oxygen is present in these films (Figure \ref{fig:RBS_ratios} and \ref{fig:aes}), which decreases resistivity by introducing charge carriers or increases resisitivity by producing interfacial oxide layers (i.e., MgO). 
Figure \ref{fig:MgWN2_resistivity} also shows that resistivity can change dramatically with composition.
Resistivity ($\rho$) increases as a function of Mg content, with Mg-poor samples exhibiting  $\rho<0.01~\Omega$-cm and Mg-rich samples exhibiting $\rho>100~\Omega$-cm.
In sum, these trends shows that the Mg-W-N system holds potential for tunable electronic properties, although future work should focus on higher quality films to bring that promise to fruition.

\section{Conclusions}
We synthesized three new polymorphs of magnesium tungsten nitrides by bulk and film synthesis methods in a previously empty ternary phase space, and demonstrated how rapid thermal annealing can be a powerful tool to reconcile thermodynamic and non-equilibrium synthesis pathways. 
Combinatorial co-sputtering yielded cation-disordered rocksalt structures across a wide composition range including \ce{MgWN2}, while samples near the \ce{Mg3WN4} stoichiometry crystallized in either a cation-disordered rocksalt or a cation-disordered hexagonal boron nitride structure. 
Rapid thermal annealing treatments of these combinatorial libraries show that rocksalt \ce{MgWN2} converts to a cation-ordered rocksaline structure at $T_\mathrm{anneal}$ = 900~\textdegree{}C, in a narrow composition window around the nominal stoichiometry.
This cation-ordered \ce{MgWN2} phase also appeared in bulk ceramic syntheses and was predicted as the ground state structure by theoretical calculations, indicating that annealing of thin film libraries can potentially access the thermodynamically stable ternary nitrides. 
Density of state calculations suggest cation-disordered rocksalt \ce{MgWN2} should exhibit metallic properties while cation-ordered rocksaline \ce{MgWN2} should exhibit semiconducting behavior.
Resistivity measurements show that rocksaline \ce{MgWN2} and rocksalt \ce{Mg3WN4} are semiconductors, but we were unable to experimentally confirm the metallic behavior of rocksalt \ce{MgWN2}. 
Resistivity varies by six orders of magnitude as a function of Mg content. 
In sum, these findings expand the toolkit through which combinatorial co-sputtering experiments can explore the thermodynamic landscape in search of new nitride compounds.

\section{Author Contributions}
C.L.R., J.R.N., and A.Z. conceptualized the project.
R.W.S. conducted GIWAXS measurements.
C.A.K. conducted bulk syntheses and analysis with support from C.L.R. and J.R.N.
K.H. conducted RBS measurements.
C.L.R. and A.Z. conducted thin film co-sputtering experiments.
A.Z. conducted annealing experiments
A.Z. and C.L.R. conducted electronic property measurements.
J.R.N. conducted DFT calculations.
C.L.R. wrote the manuscript with guidance from R.W.S., J.R.N., S.R.B., and A.Z., as well as with feedback from all other co-authors.

\section{Acknowledgements}
This work was performed in part at the National Renewable Energy Laboratory (NREL), operated by Alliance for Sustainable Energy, LLC, for the U.S. Department of Energy (DOE), under Contract No. DE-AC36-08GO28308. 
Funding provided by Office of Science (SC), Office of Basic Energy Sciences (BES), Materials Chemistry program, as a part of the Early Career Award “Kinetic Synthesis of Metastable Nitrides" (thin film studies, work conducted at NREL). 
Bulk syntheses were supported by the National Science Foundation (DMR-1653863, work conducted at Colorado State University). 
C.L.R. acknowledges support from the DOE Science Graduate Research Program (SCGSR). 
R.W.S. acknowledges support from the Director’s Fellowship within NREL’s Laboratory Directed Research and Development program. 
Use of the Stanford Synchrotron Radiation Lightsource, SLAC National Accelerator Laboratory, is supported by the U.S. Department of Energy, Office of Science, Office of Basic Energy Sciences under Contract No. DE-AC02-76SF00515. 
Thanks to Nicholas Strange for on-site support with GIWAXS measurements and to Laura Schelhas for support analyzing the data.
We thank the Analytical Resources Core at Colorado State University for instrument access and training (RRID: SCR\_021758). 
The views expressed in the article do not necessarily represent the views of the DOE or the U.S. Government.

\bibliography{MgWN2_synth.bib}

\begin{tocentry}
    \includegraphics[width=8.2cm, keepaspectratio]{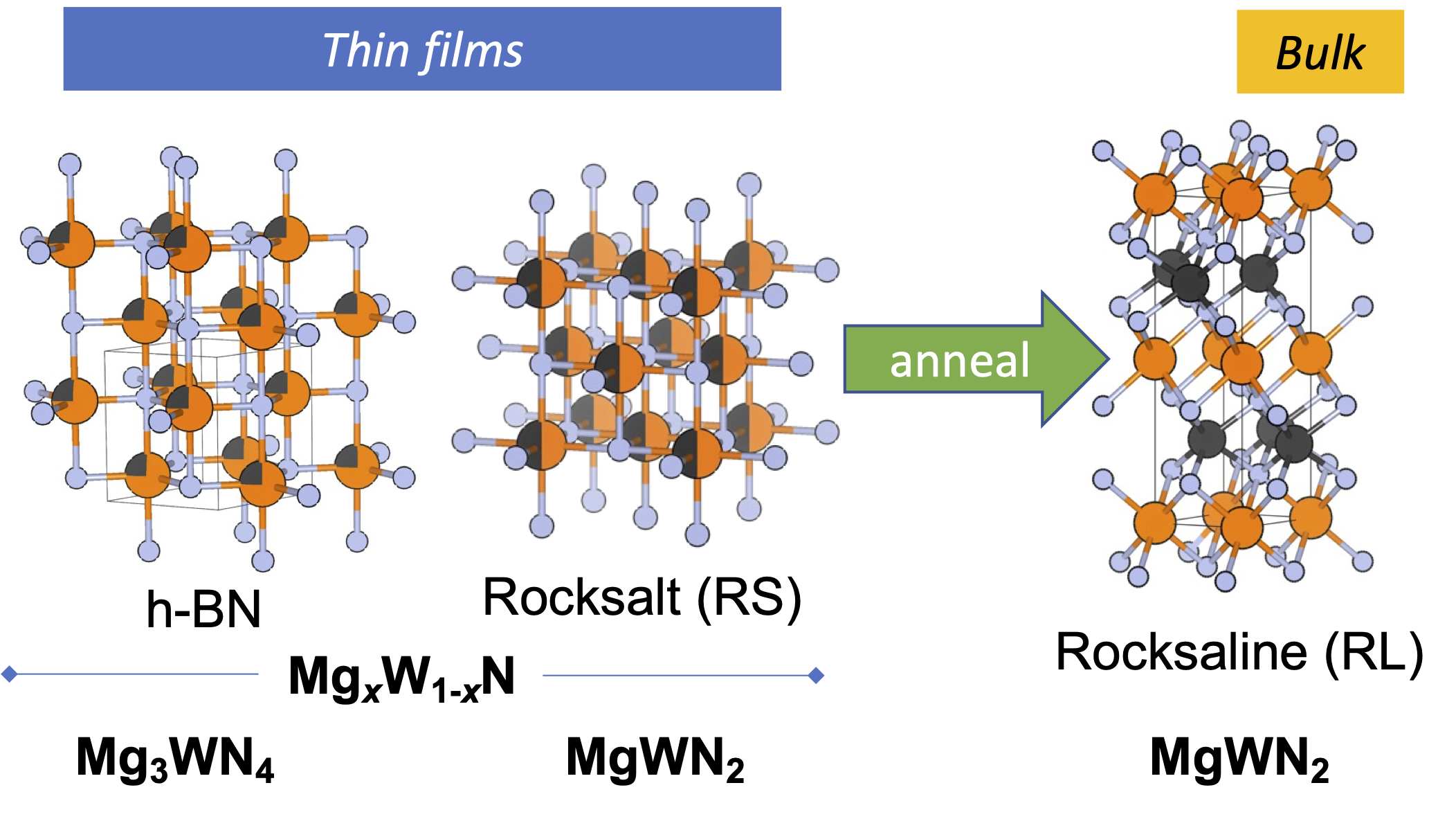}\\
\end{tocentry}

\clearpage

\section{Supporting Information for: 
Bulk and film synthesis pathways to ternary magnesium tungsten nitrides
}

\renewcommand{\thefigure}{S\arabic{figure}}
\setcounter{figure}{0}

\tableofcontents

\addcontentsline{toc}{section}{Additional experimental details}
\section{Additional experimental details}
\begin{figure}
    \centering
    \includegraphics[width = 3.2in]{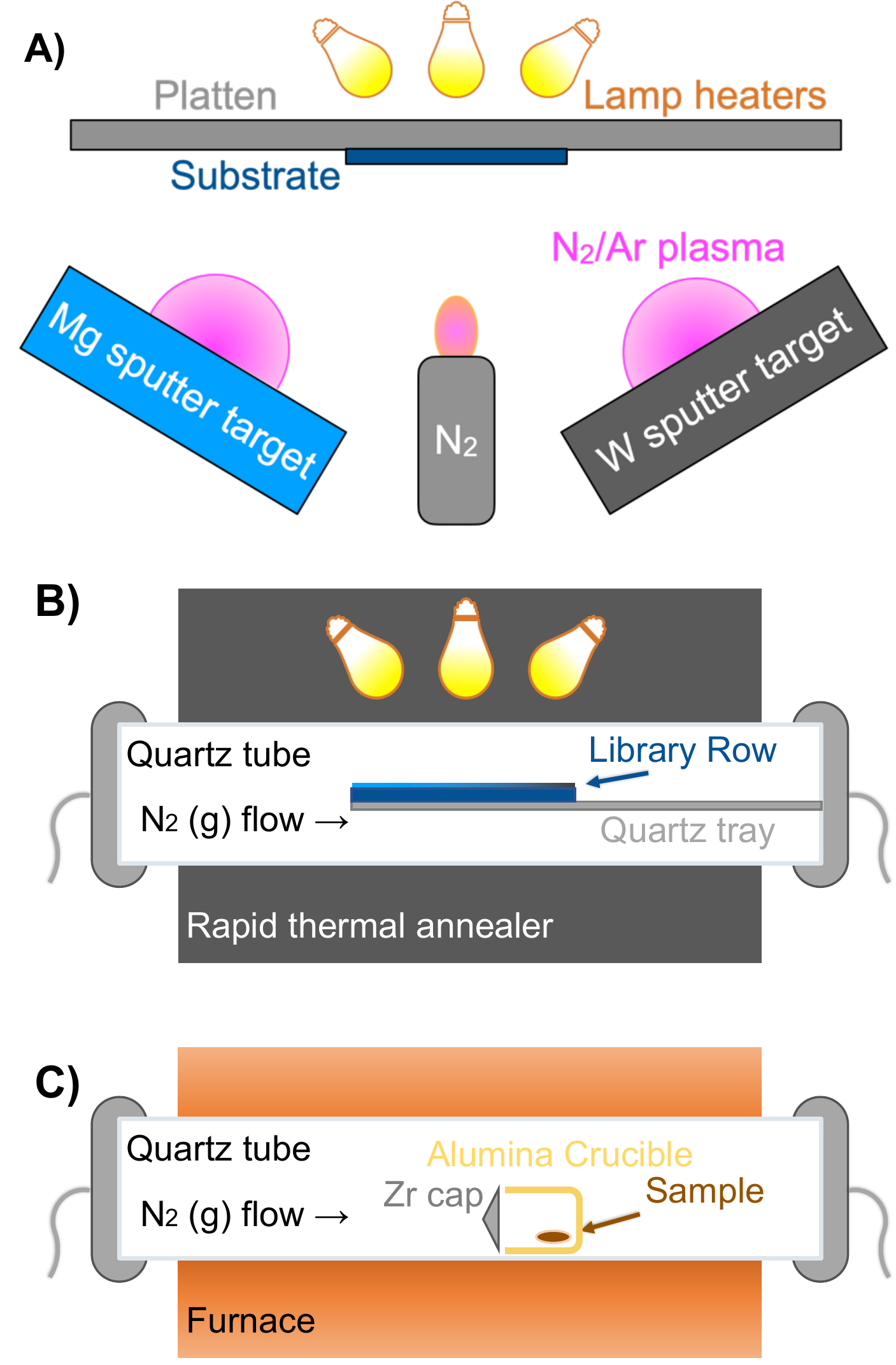}
    \caption{Three experimental setups used for the synthesis of Mg-W-N phases: A) combinatorial sputtering, B) rapid thermal annealing of combinatorial libraries, and C) bulk solid state syntheses.}
    \label{fig:experimental_setup}
\end{figure}
This manuscript combines synthetic techniques from multiple disciplines. As readers may not have personal experience with all three synthesis techniques, we diagram them in Figure \ref{fig:experimental_setup}. 
For additional information on the synthesis (and characterization) capabilities at NREL, we direct the reader to the following webpage: https://www.nrel.gov/materials-science/materials-synthesis-characterization.html

\clearpage
\addcontentsline{toc}{section}{Mg-poor phases}
\section{Mg-poor phases}
\begin{figure}
    \centering
    \includegraphics[width = 0.8\textwidth]{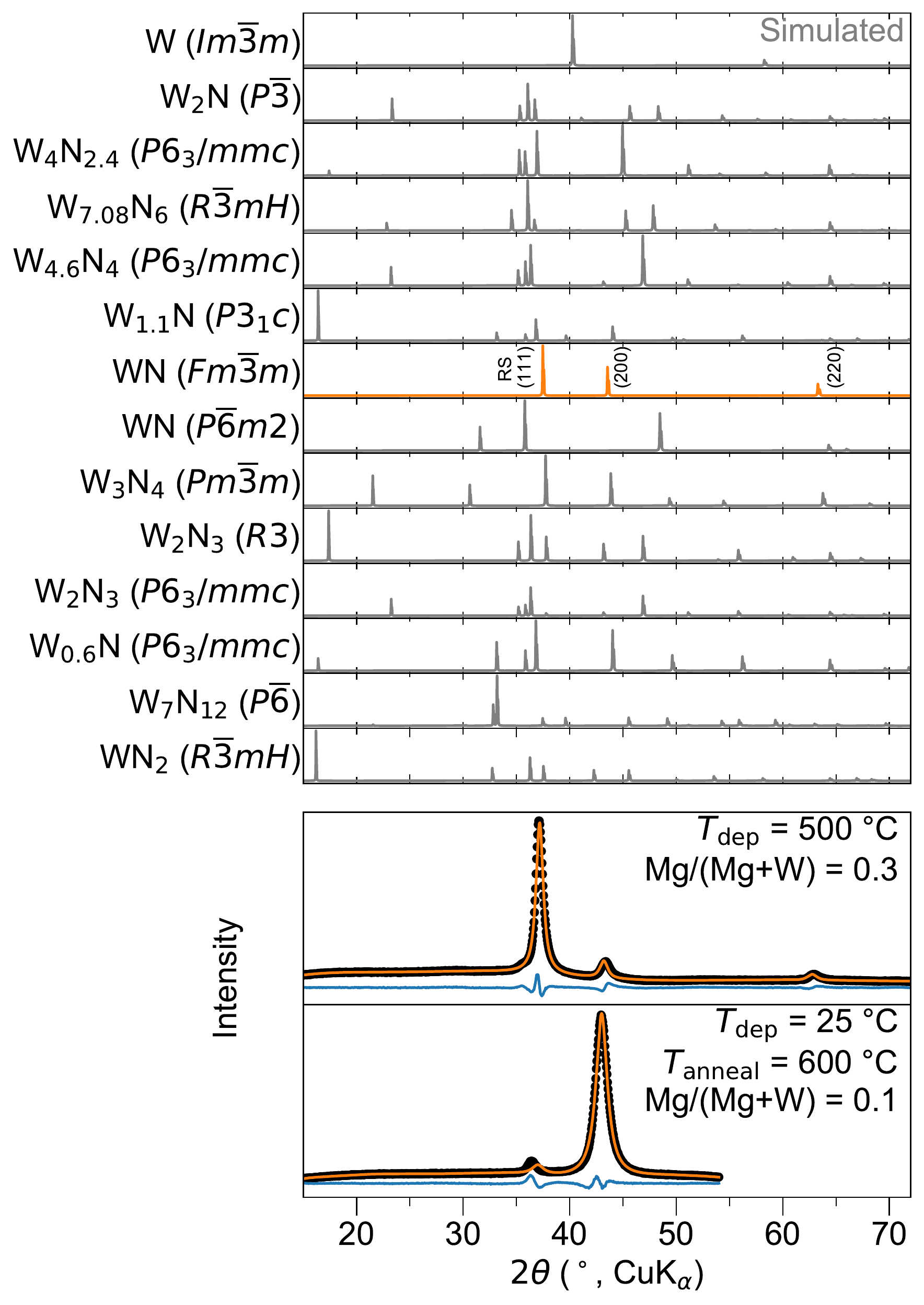}
    \caption{LeBail refinements of laboratory XRD data using the $Fm\bar{3}m$ structure for a sample deposited at 500~\textdegree{}C compared with a sample deposited at ambient conditions and annealed at 500~\textdegree{}C. Simulated patterns for WN$_x$ phases are shown for reference.}
    \label{fig:Mg_poor_phases}
\end{figure}
Mg-poor phases show slightly different structures between annealed films and films deposited at elevated substrate temperatures (Figure \ref{fig:Mg_poor_phases}). 
The $Fm\bar{3}m$ structure that best matches the Mg-poor phases from our combinatorial depositions at elevated temperatures. Therefore, we classify these compounds as the RS structure. 
However, the Mg-poor samples deposited at ambient conditions and subsequently annealed under flowing nitrogen are not well fit by the $Fm\bar{3}m$ structure. The $2\theta$ spacing between the two observed reflections is too large to match the expected the $Fm\bar{3}m$ reflections of (111) and (200) planes. 
Given the complexity of the W-N binary system and the sparcity of observed reflections, we refer to the Mg-poor phases in annealed libraries as \ce{WN$_x$} rather than the RS structure. This difference may stem from the elevated nitrogen chemical potential present in combinatorial depositions but absent during annealing. 

\clearpage
\addcontentsline{toc}{section}{Crystallographic details}
\section{Crystallographic details}
Rietveld analysis suggests some cation site disorder may exist in RL \ce{MgWN2} (Table \ref{tab:MgWN2_RL_xtal}). Refinements were performed in TOPAS v6 with the inversion parameter $x$ constrained to ensure full occupancy at both cation sites; \ce{(Mg_{1-$x$}W_{$x$})(W_{1-$x$}Mg_{$x$})N2}. 

\begin{table}[]
    \centering
    \begin{tabular}{lcccccccc}
    Wyckoff site & Atom & Multiplicity & $x$ & $y$ & $z$ & Occ. & $B (\AA{}^2$) \\ \hline
2a &Mg & 2  &0   & 0   & 0         & 0.885(10) &0.1(6) \\
2a &W  & 2  &0   & 0   & 0         & 0.115(10) &0.1(6)\\
2d &W  & 6  &1/3 & 2/3 & 0.75      & 0.885(10) &3.39(7)\\
2d &Mg & 6  &1/3 & 2/3 & 0.75      & 0.115(10) &3.39(7)\\
4f &N  & 12 &1/3 & 2/3 & 0.1453(14)& 1 & 7.4(5)\\ \hline
    \end{tabular}
    \caption{Refined atomic coordinates for RL \ce{MgWN2} in space group $P6_3/mmc$ from laboratory PXRD (Cu $K_\alpha$). Unit cell parameters were $a = 2.91097(4)$~\AA{} and $c = 10.5857(2)$~\AA{}. $R_{exp}=$ 8.40437071, $R_{wp}=$ 10.3085532, G.o.F. = 1.2265705.}
    \label{tab:MgWN2_RL_xtal}
\end{table}

\clearpage
\addcontentsline{toc}{section}{Annealing experiments}
\section{Annealing experiments}

\begin{figure}
\includegraphics[width=0.5\textwidth]{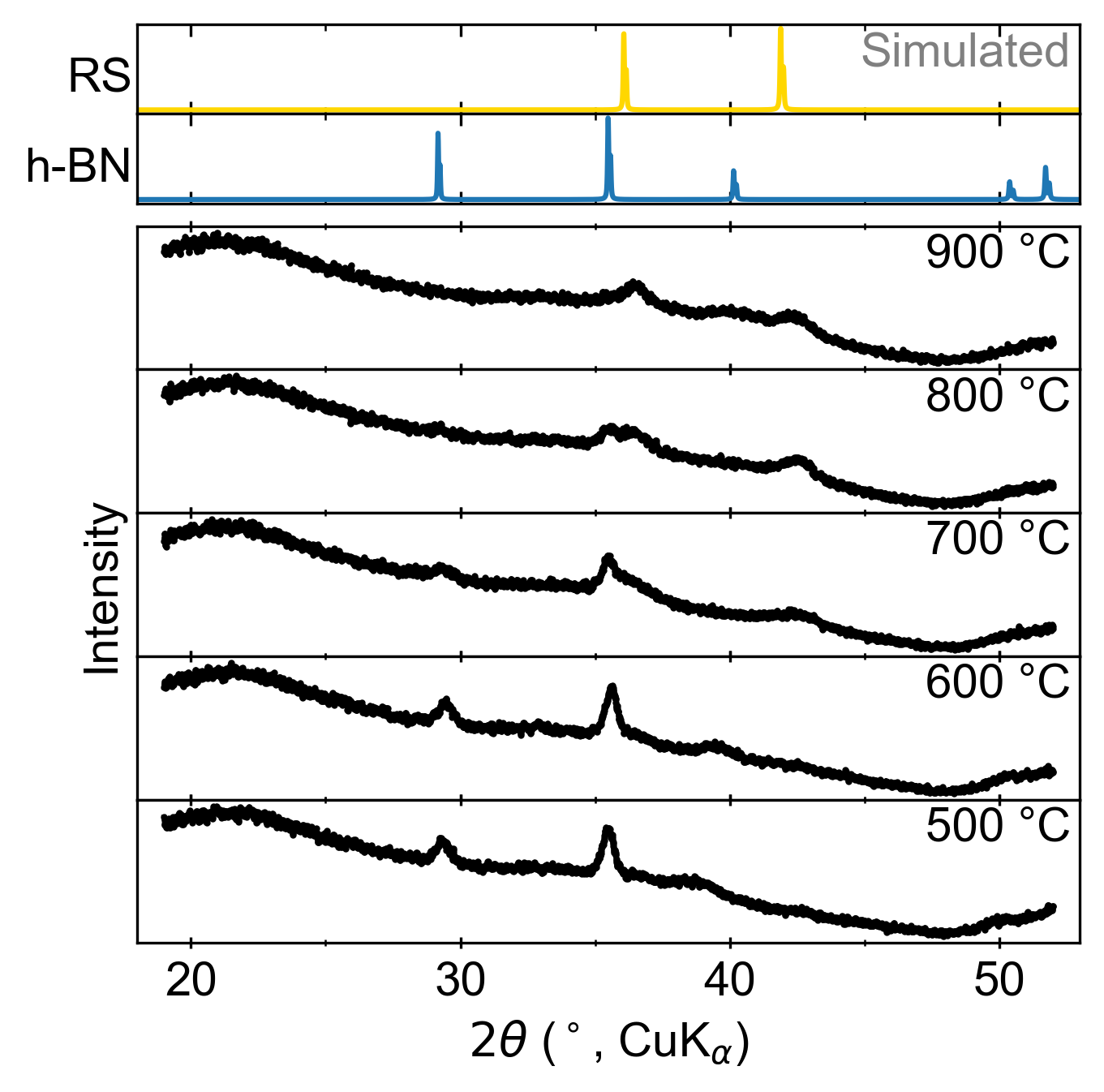}\\
  \caption{Rapid thermal annealing experiments show that the h-BN structure converts to the RS structure as $T_\mathrm{anneal}$ increases, suggesting the h-BN structure is metastable relative to the RS structure. These films were capped with a 20~nm layer of TiN to minimize Mg loss by volatility.}
  \label{fgr:MgWN_RTA_hBN_to_RS}
\end{figure}
\begin{figure}
    \centering
    \includegraphics[width = 0.5\textwidth]{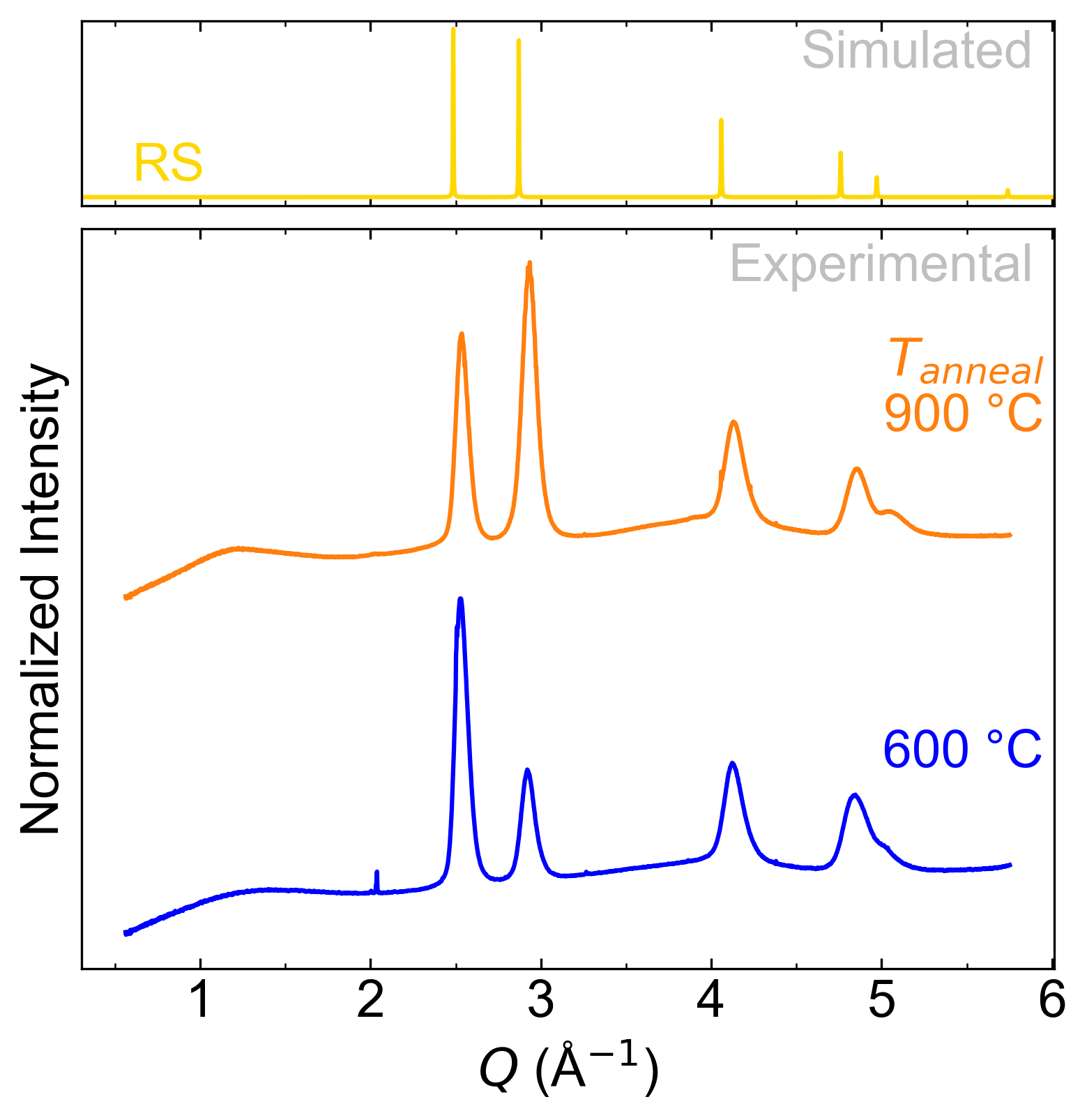}
    \caption{GIWAXS patterns for a sample of \ce{Mg3WN4} annealed at 600~\textdegree{}C and 900~\textdegree{}C show that the initial RS structure persists through these anneals. Differences in peak intensity stem from initial texturing of the deposited film, rather than changes due to annealing.}
    \label{fig:RTA_Mg3WN4_RS}
\end{figure}

RTA experiments conducted on an as-grown sample of h-BN \ce{Mg3WN4} reveal that heating this polymorph drives a conversion to the RS polymorph starting near $T_\mathrm{anneal}$=700~\textdegree{}C for 3 min, with complete conversion occurring by 900~\textdegree{}C (Figure \ref{fgr:MgWN_RTA_hBN_to_RS}). 
These films were capped with approximately 20~nm of TiN to prevent Mg loss by volatization during annealing.
This TiN layer is too thin to be detected in these XRD measurements.
This result suggests that h-BN structure of \ce{Mg3WN4} may be metastable with respect to the RS polymorph. 
In contrast, annealing the RS polymorph of \ce{Mg3WN4} does not drive any additional structural changes that are detectable by diffraction (Figure \ref{fig:RTA_Mg3WN4_RS}).

\clearpage
\addcontentsline{toc}{section}{Simulating cation disorder}
\section{Simulating cation disorder}
\begin{figure}
    \centering
    \includegraphics[width = 0.5\textwidth]{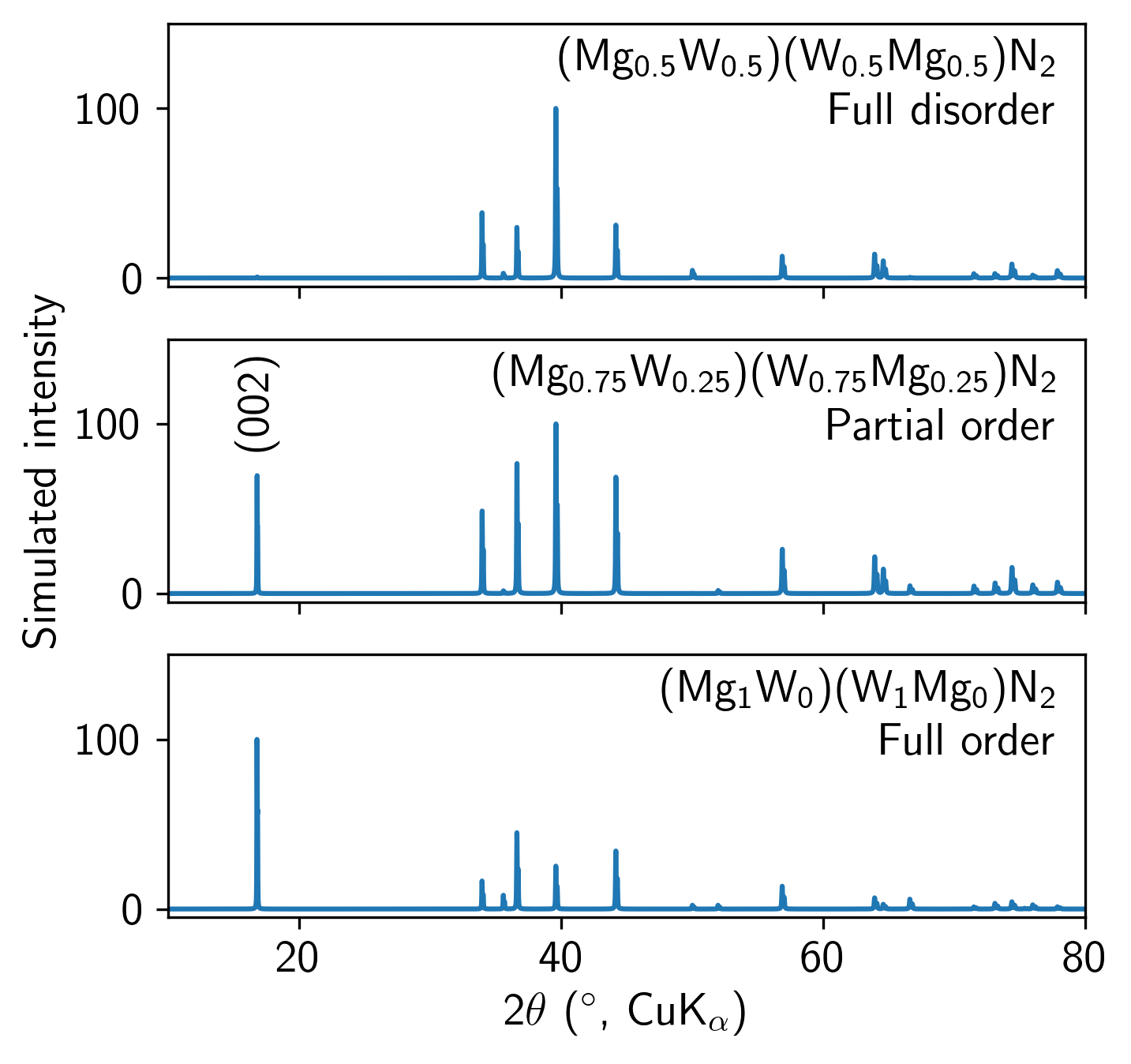}
    \caption{Simulated diffraction patterns for the \ce{MgWN2} phase in the RL polymorph ($P6_3/mmc$) with full cation disorder (top), partial cation disorder (middle), and full cation order (bottom). The relative intensity of the (002) reflection increases with increasing ordering. Patterns were simulated with VESTA software.\cite{momma2011vesta}}
    \label{fig:simulated_RL_ordering}
\end{figure}

Simulated PXRD patterns provide insight on the expected diffraction patterns for differing degrees of cation ordering. The $P6_3/mmc$ structure of the RL \ce{MgWN2} phase is primarily defined by alternating layers of Mg octahedra and W trigonal prisms (Figure \ref{fig:bulk_MgWN2}B). However, those coordination environments could persist with varying degrees of cation disorder across those two metal sites. Figure \ref{fig:simulated_RL_ordering} shows how cation disorder decreases the intensity of the (002) reflection relative to other peaks. 

\clearpage
\addcontentsline{toc}{section}{RBS measurements}
\section{RBS measurements}

\begin{figure}
    \centering
    \includegraphics[width = 0.5\textwidth]{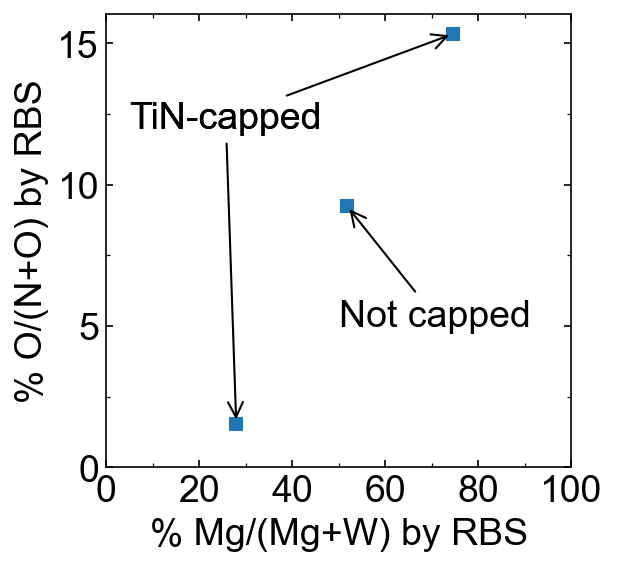}
    \caption{Oxygen as a percentage of total anions increases with increasing Mg content. Spectra are shown in Figure \ref{fig:RBS_spectra}.}
    \label{fig:RBS_ratios}
\end{figure}

Rutherford backscattering (RBS) measurements were conducted to assess N and O ratios. Select samples were deposited on a glassy carbon substrate to better resolve N and O profiles in RBS. N and O content extracted from the fits appears greater than the expected cation:anion ratio of 1:1, but we believe RBS overestimates anion content owing to difficulty in fitting such light elements when comingled with heavy W. However, as O and N are similarly affected by any systematic error, O/(N+O) ratios are still comparable sample to sample. O content increases with increasing Mg content (Figure \ref{fig:RBS_ratios}), even when a TiN capping layer is applied prior to removing the sample from the deposition chamber. This trend suggests the O may come from the Mg target or the sputtering atmosphere, rather than from reaction with air after deposition.

\begin{figure}
    \centering
    \includegraphics[width = 0.5\textwidth]{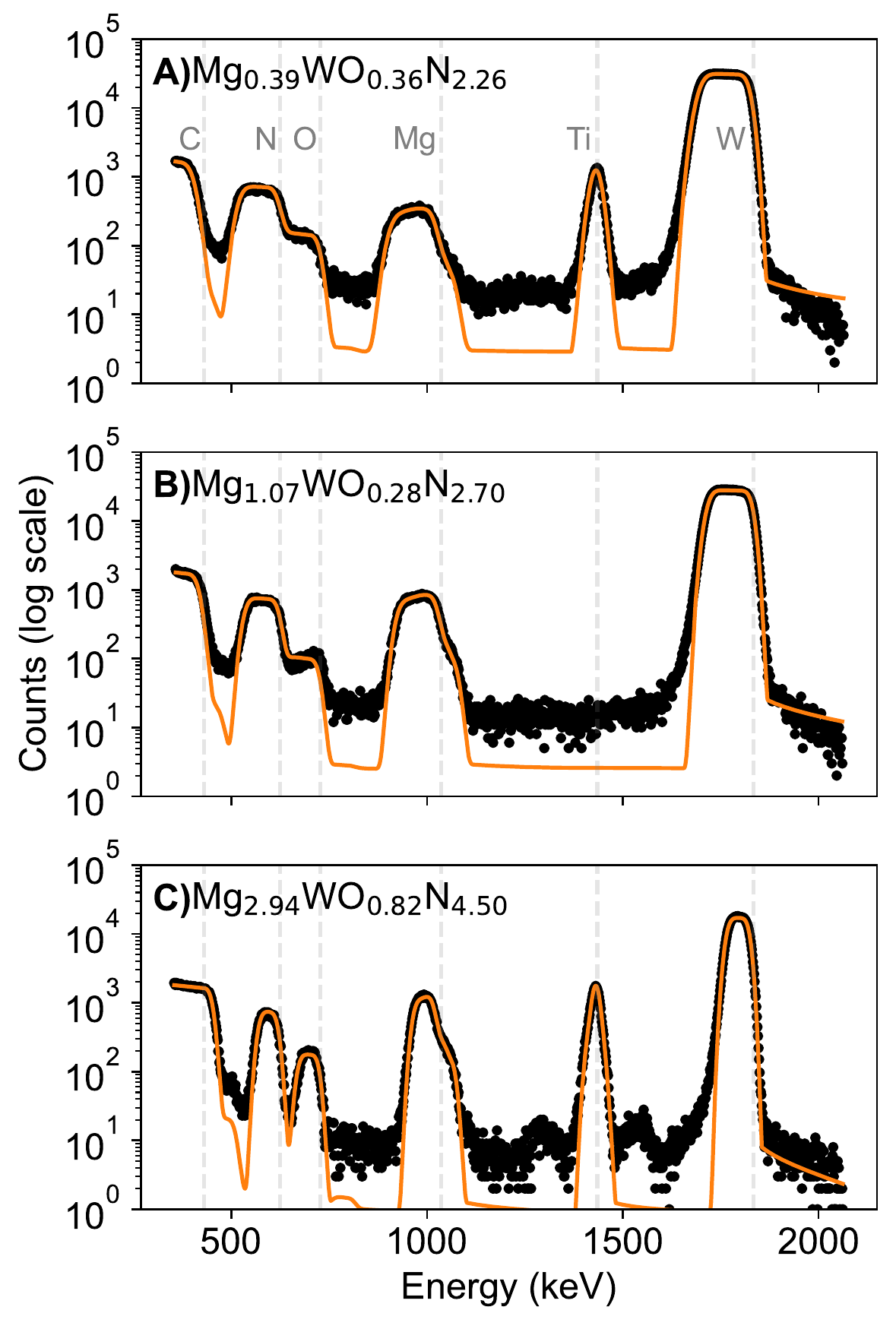}
    \caption{
    RBS spectra and fits of Mg-W-N films on a carbon substrate with A) Mg-poor RS, B) nominally \ce{MgWN2} RS and C) nominally \ce{Mg3WN4} h-BN. The films in A) and C) were capped with TiN to prevent air exposure.
    Compositions were normalized to the W signal.
    }
    \label{fig:RBS_spectra}
\end{figure}

\clearpage
\addcontentsline{toc}{section}{AES measurements}
\section{AES measurements}

\begin{figure}
    \centering
    \includegraphics[width = \textwidth]{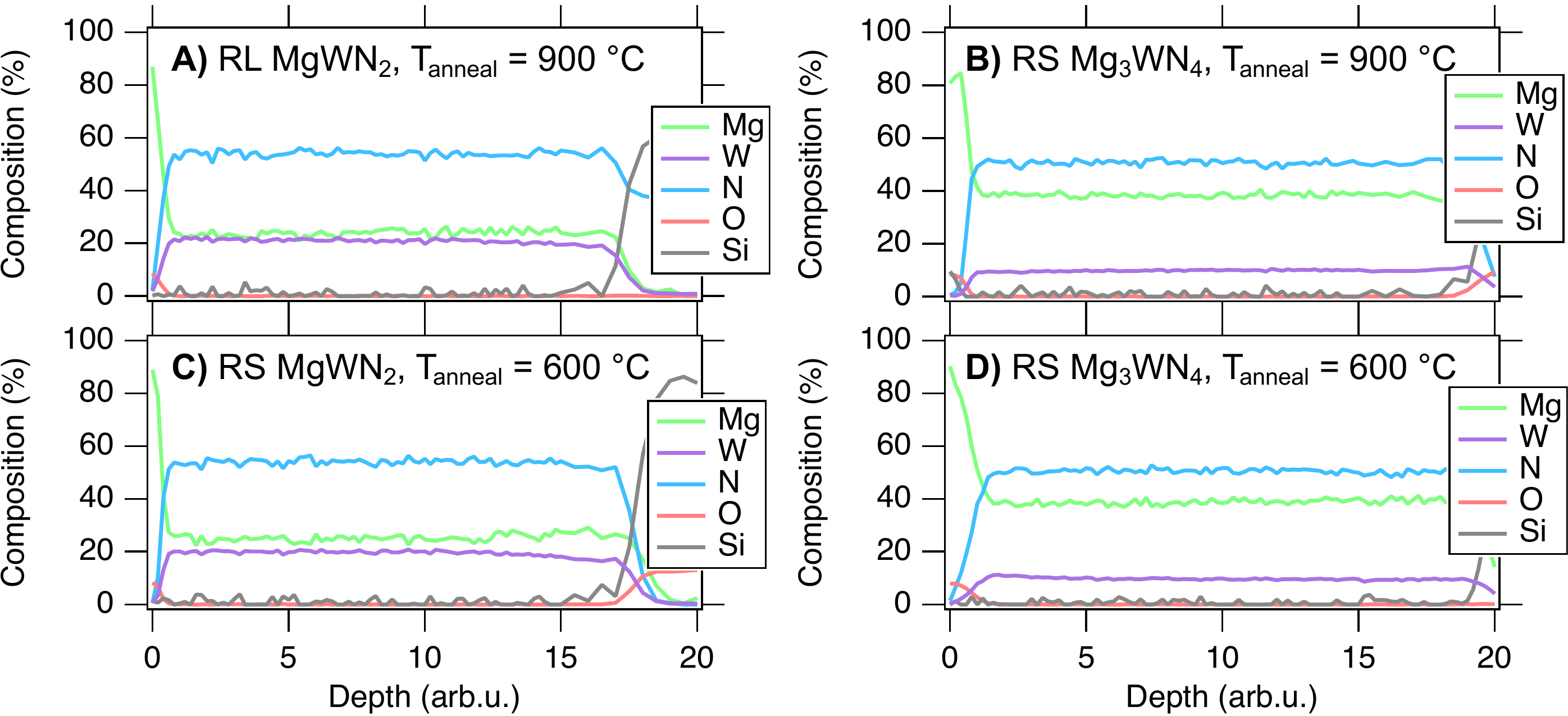}
    \caption{AES depth profiles of \ce{MgWN2} (A and C) and \ce{Mg3WN4} (B and D) films annealed at $T_\mathrm{anneal}$ = 900~\textdegree{}C (A and B) or 600~\textdegree{}C (C and D). Composition values were calibrated to the Mg/(Mg+W) ratio for \ce{MgWN2} ($T_\mathrm{anneal}$ = 900~\textdegree{}C) determined by XRF.}
    \label{fig:aes}
\end{figure}
Depth profiling by Auger electron spectroscopy (AES) was conducted on select films annealed at 600~\textdegree{}C and 900~\textdegree{}C (Figure \ref{fig:aes}). These measurements show a surface oxide layer consistent with MgO, and low oxygen levels throughout the remaining depth of the films. These oxygen levels contrast with those detected by RBS on h-BN \ce{Mg3WN4} suggesting oxide content may play a role in relative stability of the RS and h-BN structures.

\clearpage
\addcontentsline{toc}{section}{Electronic structure calculations}
\section{Electronic structure calculations}
\begin{figure}
    \centering
    \includegraphics[width = 0.5\textwidth]{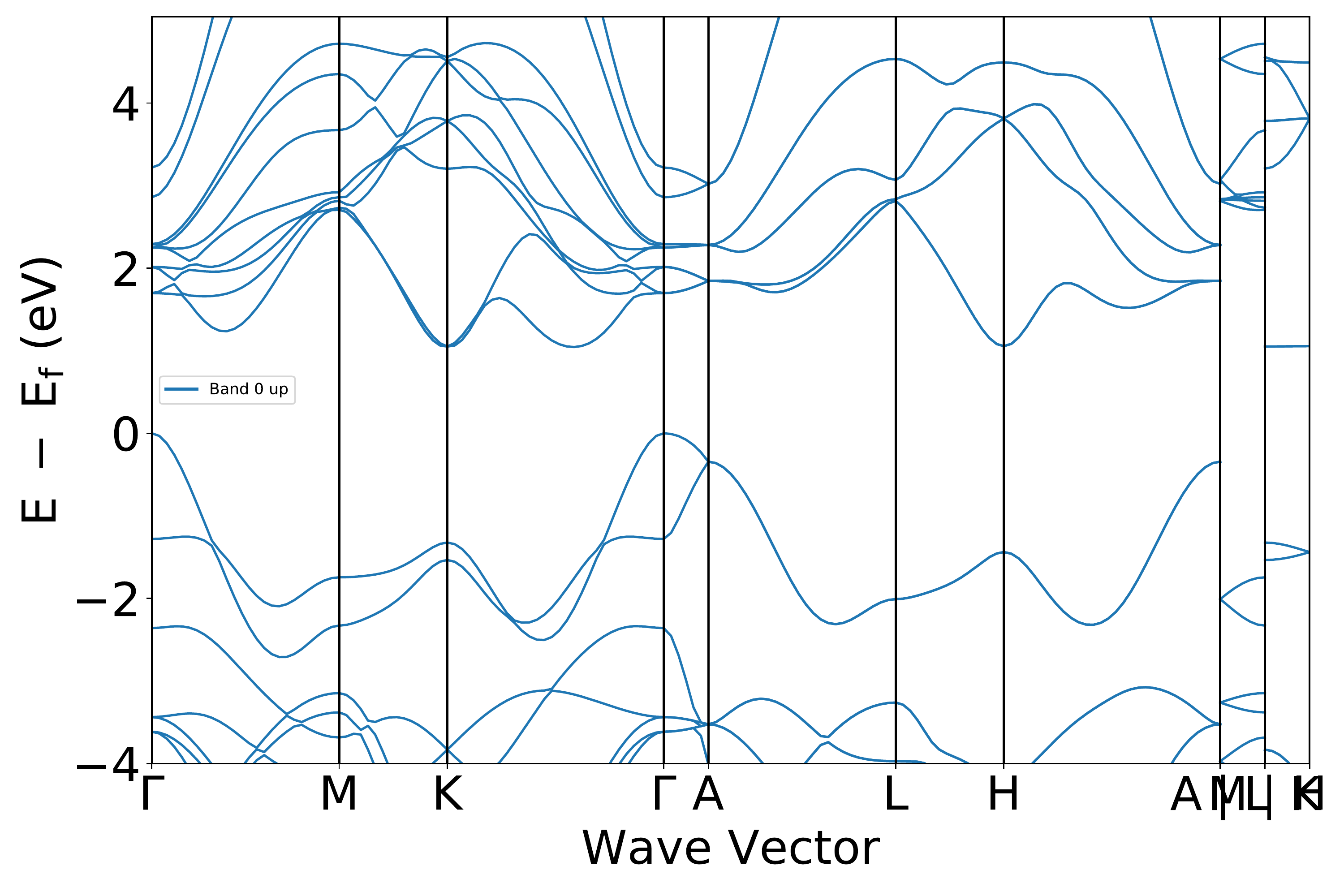}
    \caption{Calculated band structure for cation-ordered RL \ce{MgWN2} (space group $P6_3/mmc$).}
    \label{fig:bandsMgWN2_RL}
\end{figure}

\begin{figure}
    \centering
    \includegraphics[width = 0.5\textwidth]{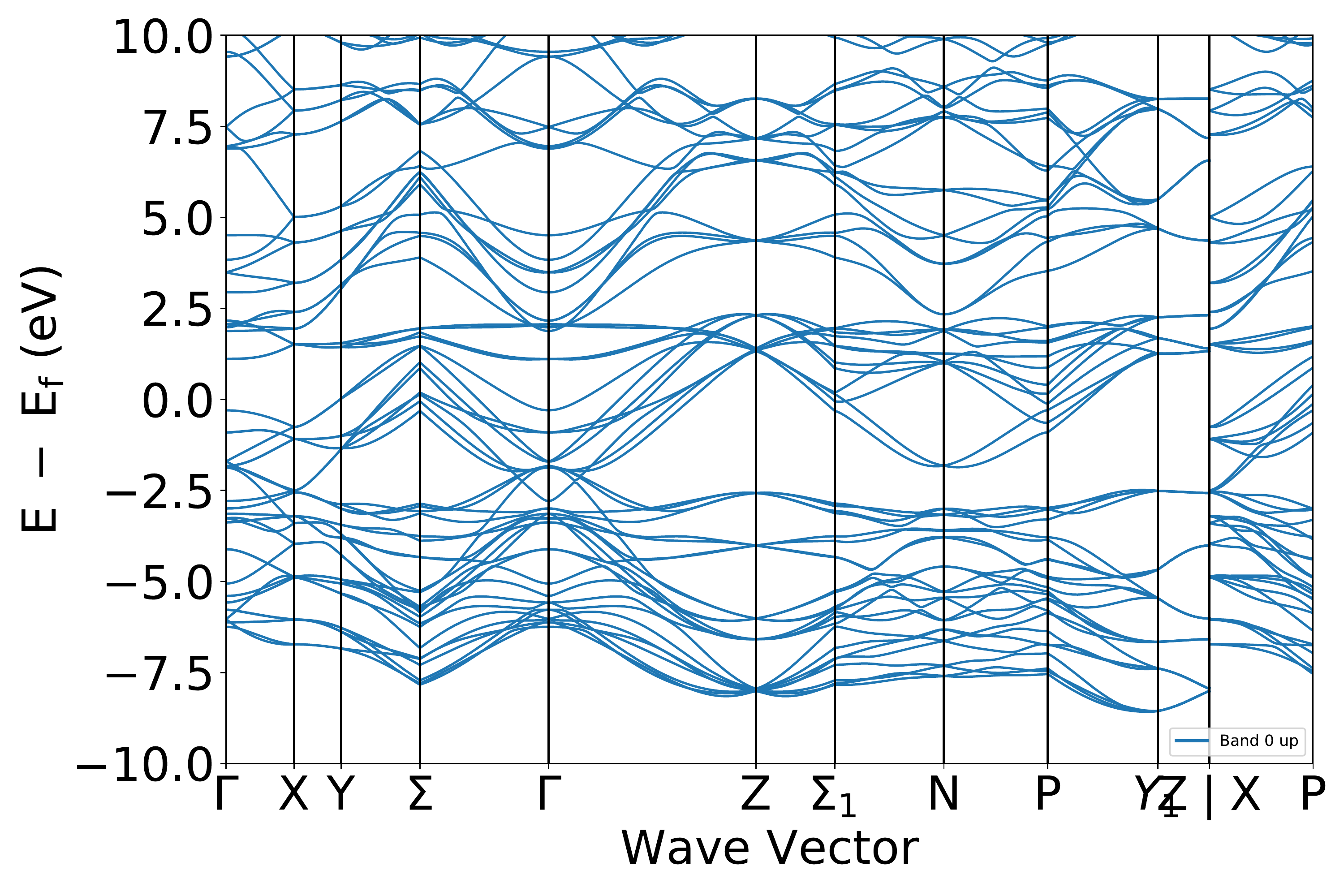}
    \caption{Calculated band structure for cation-ordered RS \ce{MgWN2} (space group $I4_1/amd$).}
    \label{fig:bandsMgWN2_RS}
\end{figure}

NSCF calculations on a uniform k-point mesh reveal a 1.18~eV indirect bandgap for RL \ce{MgWN2} (Figures \ref{fig:dos_RS_v_RL} and \ref{fig:bandsMgWN2_RL}). The valance and conduction bands are both comprised primarily of W $d$ and N $p$ states, indicating covalency in the W-N bonds. The bandgap is consistent with $d-d$ transitions expected for the $5d^2$ electron configuration of \ce{W$^{4+}$} in trigonal prismatic environment (Figure \ref{fig:dos_RS_v_RL}B).\cite{kasowski1973bandMoS2, mattheiss1973energyMoS2, verrelli_viability_2017} In contrast, the RS polymorph of \ce{MgWN2} exhibits a metallic band structure (Figure \ref{fig:bandsMgWN2_RS}), owing to the octahedral environment for \ce{W^{4+}} (Figure \ref{fig:dos_RS_v_RL}D).

\clearpage
\addcontentsline{toc}{section}{Electronic property measurements}
\section{Electronic property measurements}
\begin{figure}
    \centering
    \includegraphics[width = 0.5\textwidth]{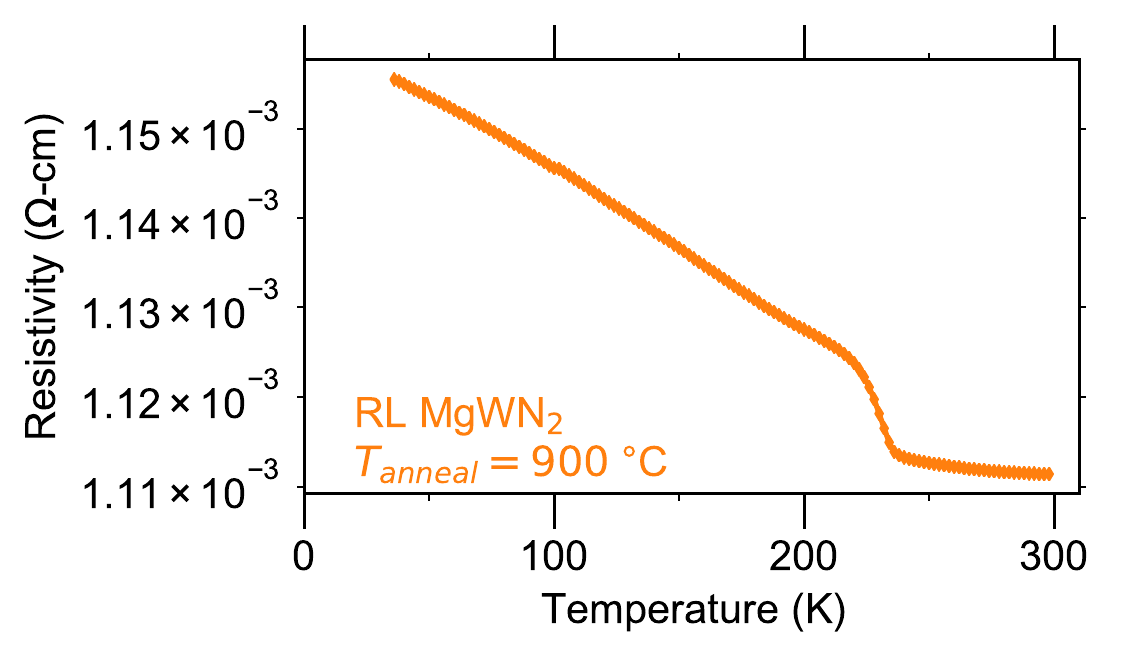}
    \caption{Temperature dependent resistivity measurements of RL \ce{MgWN2} ($T_{anneal} = 900$~\textdegree{}C) from Figure \ref{fig:MgWN2_resistivity}A, replotted on a narrower vertical axis to show the small decrease in resistivity with increasing temperature. The inflection near 230~K is an artefact of the instrument.\cite{smaha2023structuralLaWN3}}
    \label{fig:resistivity_MgWN2_RL}
\end{figure}


\end{document}